\documentclass{article}

\usepackage{PRIMEarxiv}
\usepackage{longtable,tabularx}
\usepackage[utf8]{inputenc} 
\usepackage[T1]{fontenc}    
\usepackage{hyperref}       
\usepackage{url}            
\usepackage{booktabs}       
\usepackage{amsfonts}       
\usepackage{nicefrac}       
\usepackage{microtype}      
\usepackage{lipsum}
\usepackage{graphicx}
\graphicspath{ {./images/} }
\usepackage{amsmath}
\usepackage{amssymb}
\usepackage{mathrsfs}
\title{Optimal and suboptimal spatiotemporal dynamics of multi-shear-layers in rectangular jet}

\author{
  Mitesh Thakor\thanks{\texttt{corresponding author}}\\
  Department of Mechanical and Aerospace Engineering\\
  Syracuse University\\
  Syracuse, NY 13244 \\
  \texttt{mmthakor@syr.edu} \\
   \And
  Datta V. Gaitonde \\
  Department of Mechanical and Aerospace Engineering\\
  The Ohio State University\\
  Columbus, OH 43210\\
  \texttt{gaitonde.3@osu.edu} \\
  \And
  Yiyang Sun \\
    Department of Mechanical and Aerospace Engineering\\
  Syracuse University\\
  Syracuse, NY 13244 \\
  \texttt{ysun58@syr.edu} \\
}

\begin{document}
\maketitle

\footnotetext{A part of this work was presented as Paper 2024-0285 at the AIAA SCITECH 2024 Forum, 8-12 January 2024, Orlando, FL.}

\begin{abstract}
We analyze the perturbation dynamics of a complex supersonic multi-stream rectangular jet. The dynamics are examined through application of spectral proper orthogonal decomposition (SPOD) to elicit coherent structure and linear resolvent analysis to reveal forcing-response characteristics. SPOD of a large-eddy simulation identifies Kelvin--Helmholtz coherent structures at the dominating frequency in the splitter plate shear layer region, formed by mixing core Mach~$1.6$ and bypass Mach~$1.0$ streams. Resolvent analysis leverages the time-averaged flowfield on the center plane with discounting to capture flow response over a finite time window, addressing base flow instabilities. The optimal and first sub-optimal resolvent energy amplifications peak near the dominant frequency for a wide range of frequencies and spanwise wavenumbers. Comparing the resolvent and SPOD results, we find that the linear operator over-optimizes the optimal mechanism, and the sub-optimal mode instead is more aligned with the leading SPOD mode. An intriguing \textit{shift} phenomenon where the optimal and sub-optimal gain distributions crossover is observed; these events are associated with receptive regions in the different shear layers. Subsequent input-output analyses with state-variable and spatial restrictions provide insights into componentwise amplification of the jet flow response, thus providing direction for tailored practical flow control.
\end{abstract}


\section*{Nomenclature}


{\renewcommand\arraystretch{1.0}
$M =$ mach number \\
$D_h =$   hydraulic diameter at nozzle exit \\
$\text{NPR} =$   nozzle pressure ratio \\
$\text{NTR} =$   nozzle temperature ratio \\
$Re =$   Reynolds number \\
$x =$  streamwise direction \\
$y =$  transverse direction \\
$z =$  spanwise direction \\
$\rho =$  density \\
$u =$  streamwise component of velocity \\
$v =$  transverse component of velocity \\
$w =$  spanwise component of velocity \\
$T =$  temperature \\
$P =$  pressure \\
$St =$  Strouhal number \\
$\omega  =$  radial frequency \\
$U_{\text{ref}}  =$  ambient speed of sound (reference velocity) \\
$\beta  =$  spanwise wavenumber \\
$\lambda_i  =$  $i^{th}$ eigenvalue of SPOD spectrum \\
$q  =$  vector of state variable \\
$q' =$  vector of perturbation \\
$\overline{q}  =$  temporal mean of $q$ \\
$f'  =$  non-linear terms of Navier--Stokes operator \\
$L_{\overline{q}; \beta}  =$  linearized Navier--Stokes operator \\
$H_{\overline{q}; \omega, \beta} =$  resolvent operator at given radial frequency and wavenumber \\
$\alpha =$  discounted parameter \\
$B =$  input matrix \\
$C =$  output matrix \\
$\Tilde{H}_{u\rightarrow u} =$  input-output operator indicating $u$ input to $u$ output \\
$\mathcal{U}_q =$  matrix of left singular vectors $\hat{q}_k$ (response or output modes) \\
$\mathcal{V}_f =$  matrix of right singular vectors $\hat{f}_k$  (forcing or input modes) \\
$\Sigma =$  matrix of singular values $\sigma_k$ (resolvent gain) \\
$\lambda_L =$  eigenvalues of linear operator $L_{\overline{q}; \beta}$\\
$\hat{M} =$  normalized momentum mixing based on resolvent response modes \\


\section{Introduction} \label{sec:intro}

Jet engines have evolved significantly over the last five decades,  progressing from relatively simple configurations to advanced modern designs, driven by the need for higher thrust generation, ample energy supply for advanced avionics, and the ability to meet diverse mission requirements. In the context of supersonic aerospace applications, nozzle design has become increasingly intricate, incorporating features such as fluidic inserts, geometric modifications, and thermodynamic management to meet aircraft performance criteria. Despite improved performance due to these advancements, many fundamental turbulence mechanisms responsible for noise generation remain poorly understood~\cite{goldstein1984, ashcraft2011, suder2012}. In addition to the noise signature produced by the jet~\cite{fields1993}, the unsteady turbulent flow can impose significant loading on aircraft structures and pose health hazards to flight deck crews~\cite{trost2007}, which makes mitigation of undesirable phenomena an active area of research~\cite{huff2013}.

Modern supersonic vehicles significantly reduce wave drag through advanced the integration of airframes with rectangular nozzles~\cite{bridges2012}. Modern, non-axisymmetric, airframe-integrated nozzle designs represent a significant departure from early axisymmetric geometries from design and physics perspectives, such as the growth rate of the mixing layer and major-minor axis switching of a jet plume~\cite{krothapalli1986, gutmark1989, zaman1996}.  Although offering enhanced performance flexibility, these designs come with undesirable consequences, including increased high-frequency noise generation~\cite{capone1979} and unsteady loading.

A configuration of significant interest is the three-stream, non-axisymmetric, airframe-integrated, variable-cycle engine, figure~\ref{fig:engine}(a), based on an Air Force Research Laboratory (AFRL) design. The rectangular nozzle contains multiple-streams. The merged core and bypass (fan) streams constitute the primary jet flow. The design additionally incorporates an extra bypass stream, referred to as the ``third or deck stream", to provide a cooling pad over the aircraft frame thus shielding it from the hot core flow~\cite{simmons2009}. The introduction of the third stream offers the potential to enhance efficiency from a thermodynamic perspective, contributing to reduced overall jet noise~\cite{papamoschou2001}.

The multi-stream rectangular jet flow has been investigated through experimental work at Syracuse University~\cite{magstadt2016, berry2016, magstadt2017}, complemented by numerical simulations~\cite{stack2018, stack2019} at The Ohio State University. This configuration has been shown to produce high-frequency near-field unsteady loading on the aft deck and far-field noise due to its multi-scale coherent turbulent structures (see figure~\ref{fig:engine}(b)). Passive and active control techniques have been proposed to mitigate noise and unsteady loading in non-circular jets~\cite{gutmark1999}. A passive flow control technique, specifically a wavy splitter plate trailing edge (SPTE), effectively mitigates the dominant tone of the configuration of interest~\cite{doshi2022, gist2022, kelly2024schlieren}. The wavy SPTE introduces streamwise vorticity that breaks two-dimensional coherent structures and redistributes the energy. However, passive techniques lack flexibility for different cruise and stealth scenarios. Active flow control strategies offering better control authority become crucial for such circumstances. For active flow control, the challenge lies in selecting the actuator parameters, including actuator frequency, spatial arrangement, and actuation location~\cite{cattafesta2011}.
\begin{figure} [hbpt]
    \centering
    \includegraphics[width=0.8\textwidth]{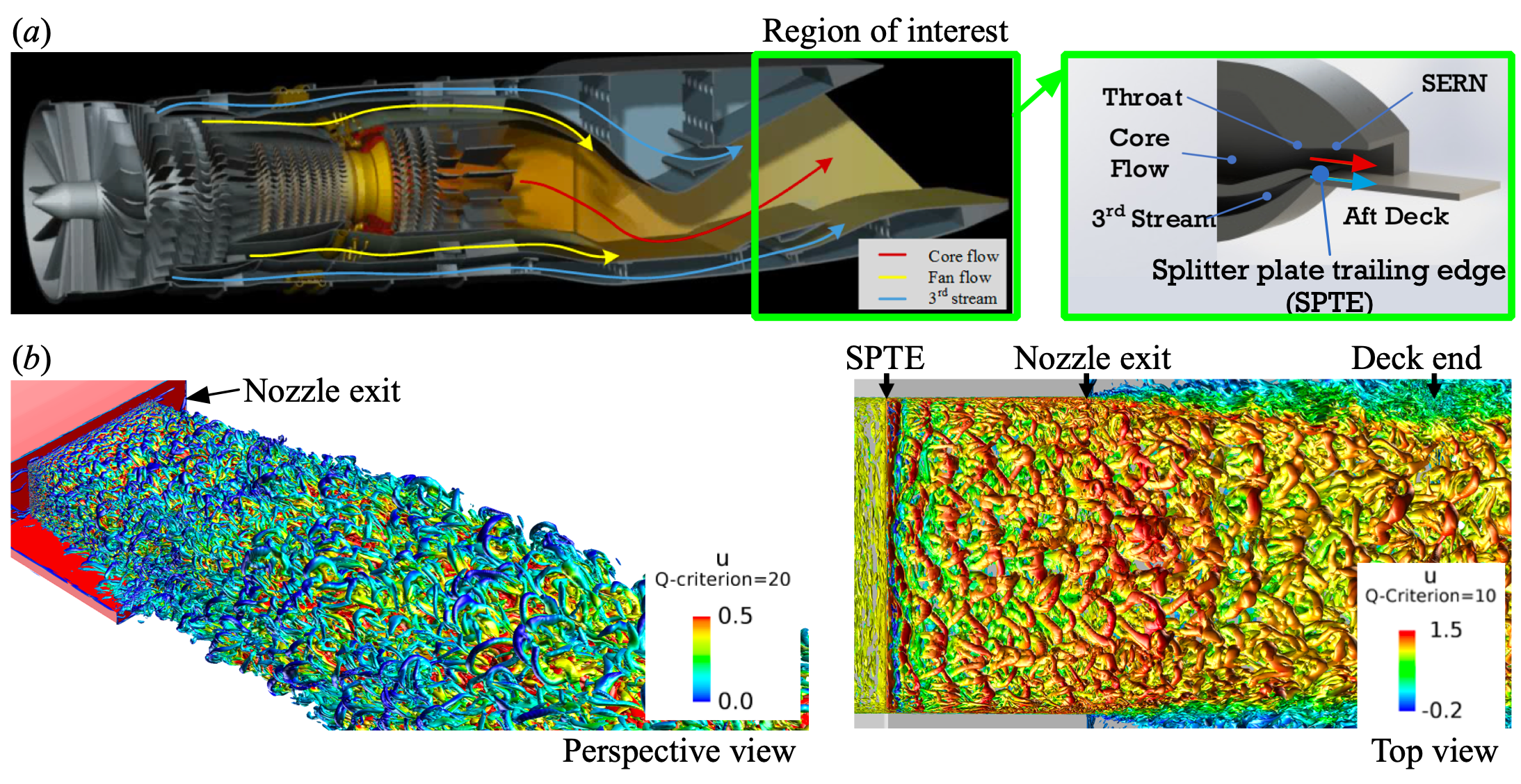}
    \caption{(a) Three-stream turbofan engine architecture developed by Air Force Research Laboratory (AFRL)~\cite{simmons2009}. The cross-section isometric view of the nozzle configuration of interest. (b) Iso-surface of Q-criterion colored by the streamwise velocity of the turbulent jet plume~\cite{stack2018}.}
    \label{fig:engine}
\end{figure}

In recent years, linear-operator-based techniques have proven efficient and effective in establishing and evaluating flow control designs~\cite{schmid2002, theofilis2011, mckeon2010, Edstrand:JFM18}. Resolvent analysis, in particular, examines amplification of harmonic forcing inputs with the linearized Navier-Stokes operator relative to a given base state, thus revealing intrinsic physics of fluid flows. Theoretically, the resolvent response modes are equivalent to the data-driven spectral proper orthogonal decomposition (SPOD) mode under a white-noise nonlinear forcing assumption~\cite{towne2018}. By analyzing the linear response to harmonic forcing, this approach has successfully explored control methods for various configurations, including flows in channels~\cite{moarref2014}, open cavities~\cite{gomez2016, sun2020, Liu:JFM21}, airfoils flows~\cite{yeh2019, yeh2020}, axisymmetric jets~\cite{garnaud2013, jeun2016, semeraro2016, towne2015}, and backward-facing steps~\cite{dergham2013}. Furthermore, resolvent forcing modes can aid in estimating optimal actuator design parameters and highlighting the most receptive regions for actuator placement. For instance, \cite{yeh2019} demonstrated its application in creating a metric to evaluate the effectiveness of spanwise actuator spacing and frequencies for reducing flow separation over an airfoil. Similarly, resolvent analysis has also guided the actuator design to mitigate noise generation in a supersonic rectangular jet flow~\cite{liang2024, prasad2024, yeung2024high}.


The present work leverages the linear-operator-based resolvent model to extract coherent structures in multi-stream rectangular jet flow. We compare these modes with the data-driven SPOD to examine the dominant amplified mechanisms identified by the linear model. Finally, we perform an input-output analysis with state restriction to identify an optimal input variable. For an isolated splitter plate shear layer, the region most receptive to perturbations is the SPTE~\cite{thakor2024}. We therefore specifically constrain the input to the SPTE in the full nozzle configuration to observe the global dynamics of the multiple shear layers.

This paper is organized as follows. Prerequisites for the current work are presented in Section~\ref{sec:method}, which summarizes the main features of the flowfield for reference (Section~\ref{sec:base_jet}) followed by short descriptions of the SPOD (Section~\ref{sec:SPODdescription} and resolvent (Section~\ref{sec:resolvent}) techniques, including discounting and the input-output formulation. Section~\ref{sec:result} presents the results starting with the SPOD spectra and coherent structures (Section~\ref{sec:SPODstructures}) and a stability analysis of the base flow (Section~\ref{sec:stability}).  Resolvent analysis without any state or spatial restrictions are divided into effects of the discounting parameter (Section~\ref{sec:dis_paramter}) and the spanwise wavenumber (Section~\ref{sec:res_finite_time}). The theoretical connections between SPOD and resolvent are leveraged in Section~\ref{sec:SPODvsResl} to aid in the interpretation of modes obtained from the two approaches. Finally, input-output analyses are deployed in Section~\ref{IO} to address effects of state  (Section~\ref{sec:io_state}) and spatial (Section~\ref{sec:io_space}) restrictions. The paper concludes with a summary of key findings in Section~\ref{sec:conclusion}.

\section{Flowfield and Methods} \label{sec:method}

\subsection{Base flow - multi-stream supersonic jet flow} \label{sec:base_jet}

Figure~\ref{fig:engine}(a) illustrates the nozzle configuration under investigation. As noted earlier, the core and fan flows of the three-stream engine are fully mixed and are designated the core (or upper) stream. The deck (or third) stream is introduced within the diverging section of the nozzle, and evolves between the splitter plate and the aft deck, which is representative of an aircraft frame and extends downstream to a length of two hydraulic diameters ($D_h$) from the nozzle exit. The rectangular cross-section at the nozzle exit has an aspect ratio of 2.7. The $D_h$ is approximately $44$ mm, and the splitter plate width measures nearly 1.85 times $D_h$. The core and deck streams are designed to operate at nozzle pressure ratios (NPR) of $\text{NPR}_{\text{core}} = 4.25$ and $\text{NPR}_{\text{deck}} = 1.89$, yielding Mach numbers of 1.6 and 1.0, respectively, based on isentropic relations. For both streams, the nozzle temperature ratio (NTR) is set to $\text{NTR}_{\text{core}} = \text{NTR}_{\text{deck}} = 1$. NPR and NTR are defined as the ratios of stagnation pressure and stagnation temperature to ambient pressure and temperature, respectively. For further details regarding nozzle dimensions and design methodology may be found in Refs.~\cite{magstadt2016, magstadt2017}.

Large-eddy simulations (LES) were performed~\cite{stack2018, stack2019} using structured finite-differences and validated against companion experiments~\cite{magstadt2016, berry2016} to confirm accuracy. The three-dimensional (3-D) computational domain and coordinate system are depicted in figures~\ref{fig:jet_mean}(a) and \ref{fig:jet_mean}(b), respectively. Throughout the manuscript, lengths are normalized by the hydraulic diameter ($D_h$). The Reynolds number is calculated as $Re_j= \rho_j \overline{u}_j D_h/\mu_j = 2.7 \times 10^6$, based on the nozzle exit conditions obtained from isentropic relations at the design Mach number and hydraulic diameter. The subscript $j$ denotes values at the nozzle exit, $\overline{u}_j$ represents the mean jet exit velocity, $\rho$ denotes density, and $\mu$ stands for viscosity. The frequency is consistently reported as Strouhal number, $St_{D_h} = \omega D_h/2\pi \overline{u}_j$, where $\omega$ is a radial frequency. The hydraulic diameter is used in normalizing frequencies, and the subscript is dropped in the following for brevity. The streamwise, transverse, and spanwise velocities are denoted by $u$, $v$, and $w$ in $x$, $y$, and $z$ directions, respectively. Pressure and temperature are denoted by $P$ and $T$, respectively.
\begin{figure} [hbpt]
     \centering
         \includegraphics[width=0.85\textwidth]{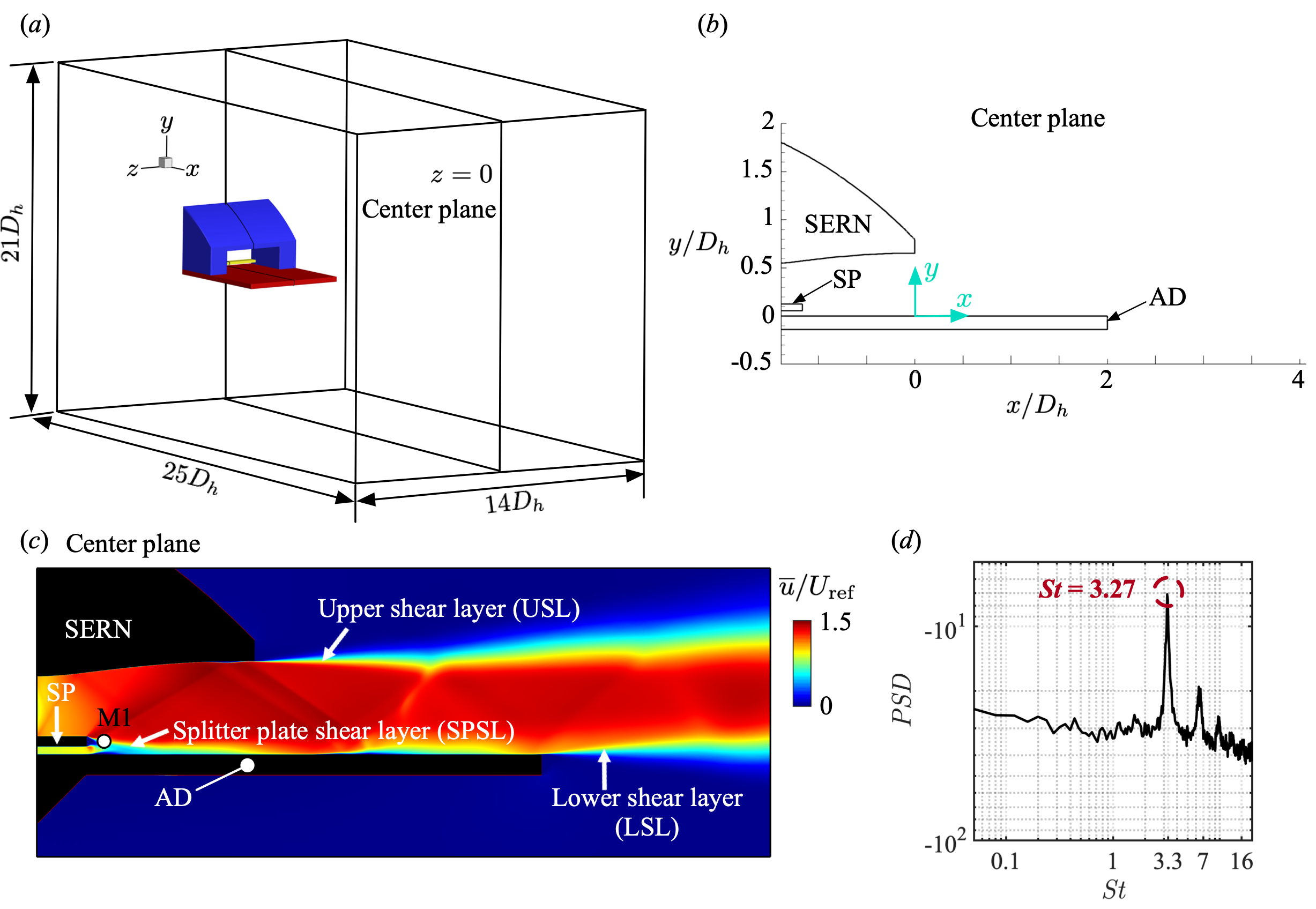}
    \caption{(a)  Schematic of 3-D LES domain (not to scale). Blue: Single-sided expansion ramp nozzle (SERN), yellow: Splitter plate (SP), and red: Aft-deck (AD). (b) Close-up view of the nozzle center plane displays the origin point of the coordinate system. (c) Time-averaged streamwise ($\overline{u}$) velocity normalized with reference velocity shows three main shear layers in the flowfield. (d) Power spectrum density (PSD) of $u-$velocity at probe M1 located in the splitter plate shear layer region, as shown in (c).}
    \label{fig:jet_mean}
\end{figure}

Figure~\ref{fig:jet_mean}(c) displays the normalized time-averaged $u$-velocity at the nozzle center plane. The ambient speed of sound ($U_{\text{ref}}$) is used for the normalization of the velocity scales. The time-averaged jet flow is derived from 6{,}845 snapshots at time intervals of $\triangle t U_{\text{ref}}/{D_h} \approx 0.016$. The convective time length for these snapshots is $t_c U_\text{ref}/ D_h \approx 110$. The statistics exhibit good convergence for this dataset. Three primary shear layers are identified from the mean flow: the splitter-plate shear layer (SPSL), resulting from the mixing of the upper core and lower deck streams; the upper shear layer (USL), formed by the mixing of the jet flow with the ambient flow on the upper side of the nozzle; and the lower shear layer (LSL), formed by a similar mixing on the lower side of the nozzle. An oblique shock is formed at the splitter plate trailing edge (SPTE), which reflects from the SERN, impinges onto the aft deck, and reflects again, creating a shock train. The power spectral density (PSD) of the streamwise velocity ($u$) for probe M1, located in the SPSL (see figure~\ref{fig:jet_mean}(c)), is depicted in figure~\ref{fig:jet_mean}(d). The probe captures the dominant vortex shedding frequency at $St = 3.27$, approximately equal to $33 \text{kHz}$. Far-field acoustic spectra capture peaks at the same frequency~\cite{magstadt2016, magstadt2017}.

\subsection{Spectral proper orthogonal decomposition} \label{sec:SPODdescription}

Spectral methods are effective tools for identifying dynamically significant coherent structures in turbulent flows~\cite{lumley1967, glauser1987, Taira_etal:AIAAJ20}. In this study, spectral proper orthogonal decomposition (SPOD)~\cite{towne2018, schmidt2020} is employed to capture and represent spatiotemporal flow statistics. To minimize the influence of nozzle sidewall effects and corner vortices, the analysis focuses on the dataset within a spanwise region of one hydraulic diameter centered at the nozzle center plane. The spanwise Fourier decomposition is performed using 72 spanwise grid points.
The spanwise wavenumbers, $\beta$, are normalized by $\beta_1$, such that $\beta/\beta_1$ represents the number of waves per width of the splitter plate. For example, $\beta/\beta_1 = 10$ indicates that ten sinusoidal waves are fitted across the width of the splitter plate. The analysis is restricted to low wavenumbers ($\beta/\beta_1 \leq 11$) to ensure convergence, as higher wavenumbers would require a more refined grid resolution.

The LES snapshots are segmented into sequence ${Q} = [{q}^{(1)}, {q}^{(2)}, ..., {q}^{(n_s)}]$. Here, $n_s$ is the total number of snapshots. The dataset is then divided into a segment or block containing $n_\text{freq}$ snapshots with an overlap. A periodic Hanning window is used over each block to prevent spectral leakage. After applying the temporal and spanwise Fourier transform, we obtain $\hat{{Q}}_{\beta, \omega_k}^{(l)} = [\hat{{q}}_{\beta, \omega_1}^{(l)}, \hat{{q}}_{\beta, \omega_2}^{(l)}, ..., \hat{{q}}_{\beta, \omega_{n_\text{freq}}}^{(l)} ] $, where $\hat{{q}}_{\beta, \omega_k}^{(l)}$ is the $l$th block realization of the transformation at the $k$th frequency ($\omega_k$) and spanwise wavenumber  ($\beta$). The cross-frequency density tensor at the given frequency and spanwise wavenumber can be given by ${S}_{\beta, \omega_k} = \hat{{Q}}_{\beta, \omega_k} \hat{{Q}}_{\beta, \omega_k}^*$. The SPOD eigenvalue problem can be solved~\cite{lumley1967} as
\begin{equation}
{S}_{\beta, \omega_k} {W} \Psi_{\beta, \omega_k} = {\Psi}_{\beta, \omega_k} {\Lambda}_{\beta, \omega_k}.
\label{eq:csd}
\end{equation}
Each column of $\Psi_{\beta, \omega_k}$ represent SPOD modes which are ranked by the eigenvalues ${\Lambda}_{\beta, \omega_k} = \text{diag}(\lambda_1, \lambda_2, ..., \lambda_N)$. Modes obtained by this method are orthonormal in the energy-norm ${W}$, in the context of Chu's energy norm~\cite{chu1965} for compressible flows defined by
\begin{equation}
E = \int \left[  \frac{R \overline{T}}{\overline{\rho}} \rho'^2 +  \overline{\rho} u_i' u_i' + \frac{R \overline{\rho}}{(\gamma-1)\overline{T}}  T'^2 \right]  \,dV.
\label{eq:E_chu}
\end{equation}

First, LES snapshots are interpolated to a uniform
grid for SPOD analysis. A rectangular domain ($-1.18 \leq x \leq 3, -0.65 \leq y \leq 1.20, -0.5 \leq z \leq 0.5$) is discretized by approximately 2.78 million grid points with state variables $[\rho', u', v', w', T']$. Based on the convergence of the modal structures and frequency resolution, 6{,}845 snapshots are segmented into 52 blocks, each block containing 256 snapshots with 50\% overlaps. This number of snapshots is sufficient since 512 snapshots per block result in similar SPOD eigenspectrum and modal structures.

\subsection{Resolvent analysis} \label{sec:resolvent}
Resolvent analysis is an operator-based method that can identify the linear intrinsic forcing and response relationships in fluid flow~\cite{farrell1993, reddy1993, schmid2002}. The compressible Navier--Stokes (NS) equations are expressed in a simplified form as,
\begin{equation}
    \frac{d { q}}{dt} = N({ q}),
    \label{eq:NS}
\end{equation}
where the state variable vector is $ q = [\rho, u, v, w, T] \in \mathbb{R}^{5n}$. The number of grid points used to discretize the computational domain is denoted by $n$. We decompose the full flow state ${q}(x,y,z,t)$ into a two-dimensional base state (time-averaged or mean) $\overline{ q}(x,y)$ and 3-D unsteady perturbation ${q'}(x,y,z,t)$ as follows,
\begin{equation}
     q (x,y,z,t) =  \overline{ q} (x,y) +  q' (x,y,z,t).
    \label{eq:decom}
\end{equation}
After substituting this state decomposition in Eq.~(\ref{eq:NS}) and subtracting mean flow equation, we linearize the NS equation and the perturbation equation as,
\begin{equation}
    \frac{d { q'}}{dt} = \Tilde{L}_{\overline{ q}}{ q'} +  f',
    \label{eq:per}
\end{equation}
where $\Tilde{L}_{\overline{ q}}$ represents the linearized Navier-Stokes operator, and $ f'$ denotes the non-linear terms or external forcing applied to the linear system. For laminar flows, $\overline{ q}$ is a steady state~\cite{trefethen1993, jovanovic2004}. Whereas for turbulent flows, $\overline{ q}$ can be a non-equilibrium but statistically stationary mean flow and the forcing term  $ f'$ can be interpreted as a combination of nonlinear terms and the external forcing applied to the fluid flow system~\cite{mckeon2010}. Next, the forcing $ f'$ and perturbation $ q'$ are expressed as the Fourier representation in the bi-global perturbation framework of
\begin{equation}
     {f'}(x,y,z,t) =  {\hat{f}}_{\omega, \beta}(x,y) e^{i(\beta z - \omega t)},
    \label{eq:four_f}
\end{equation}
\begin{equation}
     {q'}(x,y,z,t) =  {\hat{q}}_{\omega, \beta}(x,y) e^{i(\beta z - \omega t)},
    \label{eq:four_q}
\end{equation}
where $\omega$ and $\beta$ are real-valued radian frequency and spanwise wavenumber, respectively. The spatial amplitude functions for the forcing and perturbation are given by $ {\hat{f}}_{\omega, \beta}$ and $ {\hat{q}}_{\omega, \beta}$, respectively. After substituting Eqs.~(\ref{eq:four_f}) and (\ref{eq:four_q}) into Eq.~(\ref{eq:per}), the linearized governing equation can be written as,
\begin{equation}
     {\hat{q}}_{\omega, \beta} = [-i \omega I - L_{\overline{ q}; \beta }]^{-1}  {\hat{f}}_{\omega, \beta} = {H}_{\overline{ q}; \omega, \beta}  {\hat{f}}_{\omega, \beta},
    \label{eq:transfer_H}
\end{equation}
where, ${H}_{\overline{ q}; \omega, \beta} = [-i \omega I - L_{\overline{ q}; \beta }]^{-1}$ referred as a {\it resolvent operator} is a transfer function between forcing ${\hat{ f}}_{\omega, \beta} $ and flow response ${\hat{ q}}_{\omega, \beta}$ for the given base state $\overline{ q}$ at the prescribed spanwise wavenumber $\beta$ and radian frequency $\omega$.

\subsubsection{Discounted window} \label{sec:discount}

Resolvent analysis is typically performed by sweeping the real-valued frequency ($\omega$) component of the Laplace variable, defined as $s = \omega + i\alpha$. For stable flows, it is common to select a real-valued $\alpha$ equal to zero, effectively reducing the Laplace transform to the Fourier transform~\cite{mckeon2010, luhar2015}. In such cases, where all eigenvalues of the stable operator $L_{{\overline{q}}}$ lie below the imaginary axis, the Fourier transform is an appropriate choice. However, a statistically stationary time-averaged flow at a high Reynolds number does not guarantee stable $L_{{\overline{q}}}$. In such instances, resolvent analysis should be conducted with a shifted imaginary axis (i.e., $\alpha > 0$) to ensure that all eigenvalues are positioned below the positive imaginary axis~\cite{sun2020, yeh2019}. The parameter $\alpha$ must satisfy the condition $\alpha > \text{max} (\text{Im}(\lambda_L))$, where $\lambda_L$ represents the eigenvalues of $L_{{\overline{q}}}$. This modification is equivalent to the exponential discounted approach introduced by~\cite{jovanovic2004}.
The discounted resolvent operator can be formulated as
\begin{equation}
 H^{\alpha}_{\overline{ q}; \omega, \beta} = [-i (\omega + i \alpha) I - L_{\overline{ q}; \beta }]^{-1}.
    \label{eq:H_disc}
\end{equation}
The discounted resolvent operator Eq.~(\ref{eq:H_disc}) can be equivalently expressed as $H^{\alpha}_{\overline{ q}; \omega, \beta} = [-i \omega I - (L_{\overline{ q}; \beta } - \alpha I)]^{-1}$, which is a transfer function for the harmonic forcing to the response through the stable operator $(L_{\overline{ q}; \beta } - \alpha I)$. Furthermore, this discounted modification can be viewed as a finite-time horizon input-output analysis, where a finite time window is characterized by $1/\alpha$ \cite{yeh2019, yeh2020, sun2020}. In other words, a larger value of $\alpha$ corresponds to a shorter time window within which disturbances are allowed to grow.

The discounted analysis provides a means to extract local flow physics from a global resolvent operator by selecting an appropriate range of $\alpha$ (the discounted parameter). This approach enhances the utility of the analysis for flow control applications by offering insights into localized flow dynamics.

\subsubsection{Input-output formulation} \label{sec:input-output}

The linear perturbation Eq.~(\ref{eq:per}) can be rewritten by introducing an input matrix $B$ and an output matrix $C$ as
\begin{equation}
    \frac{d { q'}}{dt} = \Tilde{L}_{\overline{ q}}{ q'} + B  f',
    \label{eq:per_IO}
\end{equation}
\begin{equation}
    g = C  {q'},
    \label{eq:per_output}
\end{equation}
where $g$ is an output or measurement built upon the state $ q'$ using output matrix $C$. Term $B  f'$ prescribes an input of the system. The input and output can be tailored by manipulating the matrices $B$ and $C$. This type of analysis is referred to as a componentwise input-output analysis \cite{jovanovic2005}. The above system of equations yields a transfer function between input and output in a Fourier domain for a given frequency $\omega$ and wavenumber $\beta$ as,
\begin{equation}
\Tilde{  {H}}_{\overline{ q}; \omega, \beta} = C [-i \omega I - L_{\overline{ q}; \beta }]^{-1} B.
\end{equation}
Here, we use $\Tilde{  {H}}$ to represent the input-output transfer function to differentiate it from the classical resolvent transfer function ${H}$.

We can restrict the state variables and spatial location of the input and output by manipulating $B$ and $C$ matrices in $\Tilde{  {H}}_{\overline{ q}; \omega, \beta} = C [-i \omega I - L_{\overline{ q}; \beta }]^{-1} B$. For instance, a streamwise velocity output is observed for a streamwise velocity input, then $B$ and $C$ are given by
\begin{gather}
B = C =
\begin{bmatrix}
0 & 0 & 0 & 0 & 0  \\
0 & I & 0 & 0 & 0  \\
0 & 0 & 0 & 0 & 0  \\
0 & 0 & 0 & 0 & 0  \\
0 & 0 & 0 & 0 & 0 \\
\end{bmatrix} ,
\label{c_out}
\end{gather}
where, $I$ is an identity matrix of size $n \times n$.

The resolvent analysis or input/output can be cast in the framework of singular value decomposition (SVD) of the resolvent operator to determine the forcing $\hat{f}_{\omega,\beta}$, the response $\hat{q}_{\omega,\beta}$, and their gain (i.e., energy amplification). The SVD decomposition of the resolvent operator is given by,
\begin{equation}
{H}_{\overline{ q}; \omega, \beta} = \mathcal{U}_q \Sigma \mathcal{V}_f^*,
\end{equation}
where, $\mathcal{U}_q = [\hat{q}_1,\hat{q}_2,...,\hat{q}_k]$ is a set of left singular vectors $\hat{q}_j$ called the response (output) modes, and $\mathcal{V}_f = [\hat{f}_1,\hat{f}_2,...,\hat{f}_k]$ contains a set of right singular vectors $\hat{f}_j$ called the forcing (input) modes, with the superscript $^*$ representing the Hermitian transpose. The matrix $\Sigma = \text{diag}(\sigma_1,\sigma_2,...,\sigma_k)$ gives the amplification ratios of the response and forcing modes based on a selected norm. The singular values are in descending order $(\sigma_1\geq\sigma_2\geq...\geq\sigma_k)$, and the first singular value $\sigma_1$ is referred to as the optimal resolvent gain. A rank-1 assumption can often be made if $\sigma_1 \gg \sigma_2$, meaning the physical mechanism associated with the optimal (first) forcing-response pair dominates flow physics~\cite{beneddine2016, towne2018}. Moreover, the above expression can also be rewritten in terms of $H \mathcal{V}_f = \mathcal{U}_q \Sigma$, which is interpreted as that each column of $\mathcal{V}_f$ is a forcing vector that is mapped to the corresponding column of a response $\mathcal{U}_q$ through the transfer function ${H}$~\cite{schmid2002}.

The resolvent gain is studied in the context of Chu's energy norm for compressible flows as given by Eq.~(\ref{eq:E_chu}) by performing SVD on the weighted resolvent matrix $W^{1/2} {H}_{\overline{ q}; \omega, \beta} W^{-1/2}$. The weight matrix $W$ can be constructed based on the discretization scheme adopted in the numerical configuration. Then, the resulting left and right singular vectors should be scaled by $W^{-1/2}$ to obtain the correct response and forcing modes, respectively.


\section{Results} \label{sec:result}


\subsection{SPOD spectra and coherent structures} \label{sec:SPODstructures}

Figure~\ref{fig:spod_spectra} presents the SPOD eigenvalue spectrum for representative spanwise wavenumbers.
A significant separation between the first and second eigenvalues is observed in the frequency range $0.2 \leq St \leq 1$, as well as at the dominant frequency $St=3.27$  and its first two harmonics for $\beta/\beta_1 = 0$ as shown in figure~\ref{fig:spod_spectra}(a). When the first mode is substantially more energetic than the second mode (i.e., $\lambda_1 \gg \lambda_2$), the flow demonstrates rank-1 behavior, with the physical mechanism associated with the first mode being dominant. The rank-1 behavior diminishes for non-zero spanwise wavenumber. For $\beta/\beta \approx 5$ and 11, small peaks are observed in the first eigenvalue at $St=3.27$ and its first two harmonics. However, these peaks contain significantly less energy compared to $\beta/\beta = 0$. This suggests that the physical mechanism associated with the dominant frequency $St=3.27$ is predominantly two-dimensional (2-D). This observation is consistent with the findings of~\cite{stack2019}, who identified 2-D pressure-coherent structures in the SPSL region from a three-dimensional dataset.
\begin{figure}[hbpt]
    \centering
         \includegraphics[width=0.85\textwidth]{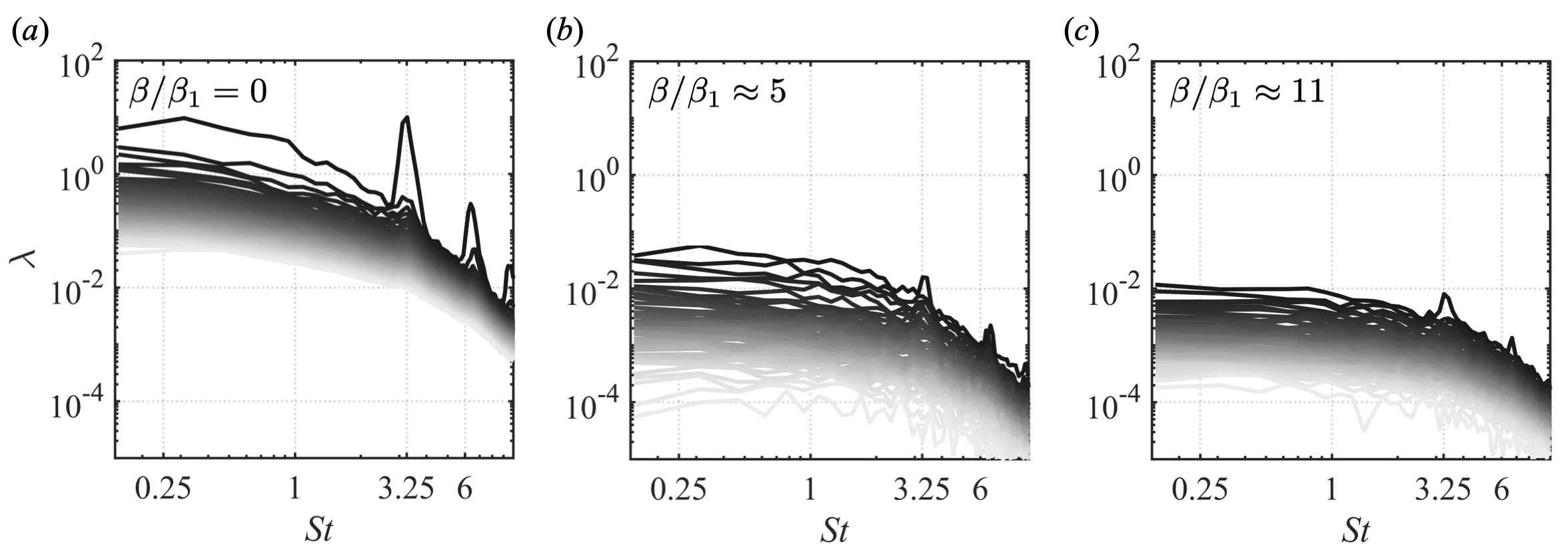}
    \caption{SPOD eigenvalue spectra for the spanwise wavenumber $\beta/\beta_1$ = (a) 0, (b) 5, and (c) 11. At each frequency, decreasing eigenvalues are shown in lighter shades, i.e., $\lambda_1 \geq \lambda_2 ... \geq \lambda_N$.}
    \label{fig:spod_spectra}
\end{figure}

The first three SPOD modes of the $v$-velocity are shown in figure~\ref{fig:spod_modes} for two spanwise wavenumbers at representative frequencies. At the dominant frequency ($St = 3.27$), where the flow is rank-1, the leading 2-D ($\beta/\beta_1 = 0$) wavepackets appear in the SPSL region. These modal structures resemble the Kelvin--Helmholtz (KH) instability pattern of the base flow. Strong acoustic waves originating from the SPTE and traveling to the far field are also observed. Meanwhile, Modes~2 and~3 display wavepackets in all three shear-layer regions, i.e., SPSL, LSL, and USL; in this case, $\lambda_2$ and $\lambda_3$ are not well separated. That is, the wavepackets appear in a single shear layer when eigenvalues are well separated from each other. The SPOD leading modes indicate that the USL dynamics are associated with low frequencies. For instance, the leading wavepackets are in the USL for low frequencies ($St = 0.62$, 2.1), and the sub-dominant modes are in multi-shear-layer regions. In contrast, the wavepackets are present in all three shear layer regions at the higher frequency $St = 5.75$ for the first three modes, where $\lambda_1 \approx \lambda_2 \approx \lambda_3$.
\begin{figure}[hbpt]
     \centering
         \includegraphics[width=1\textwidth]{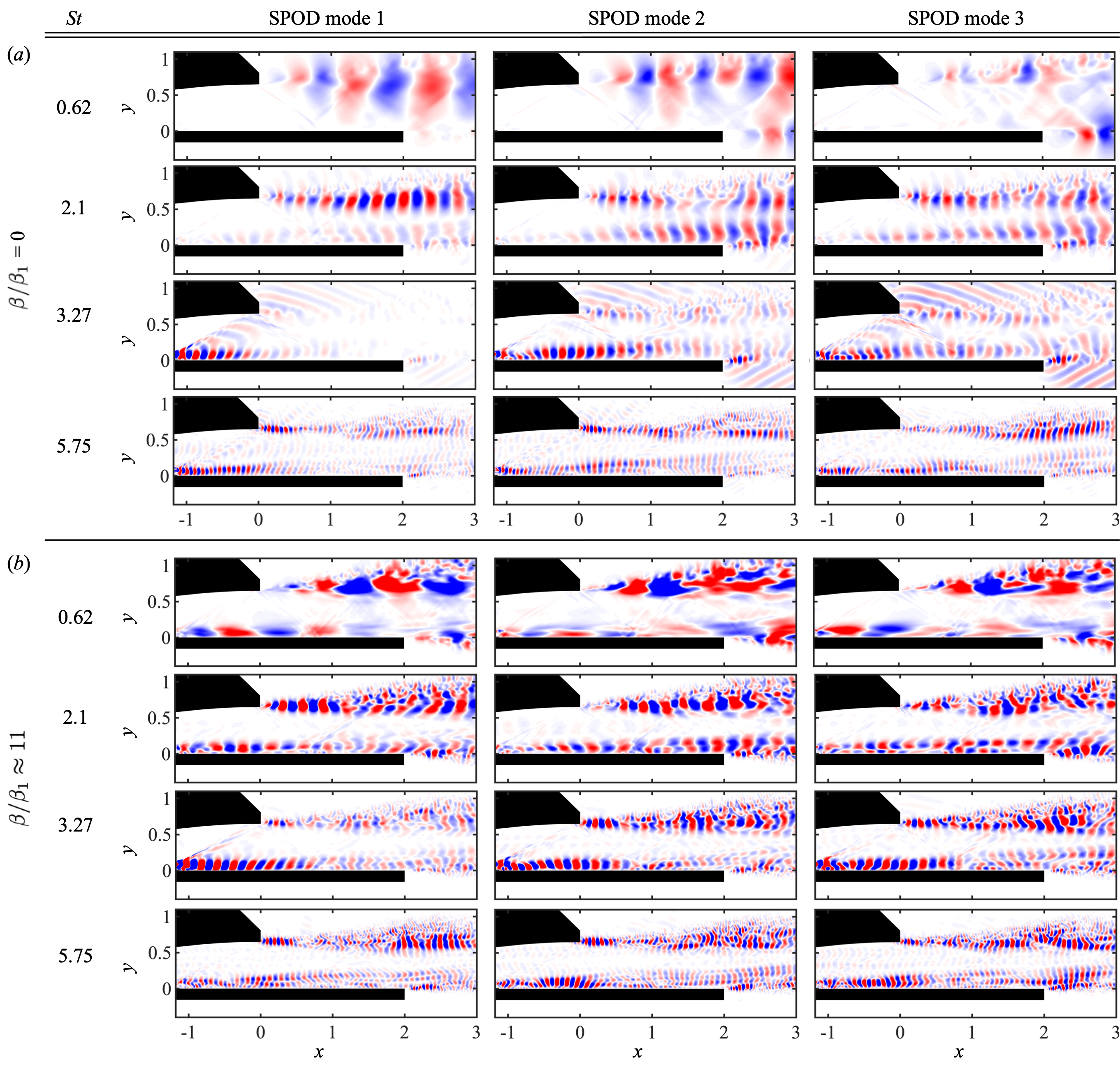}
    \caption{First three modes (real component of $v$-velocity) of SPOD analysis for the spanwise wavenumber $\beta/\beta_1 =$ (a) 0 and (b) 11 at representative frequencies. [blue, red] = [-0.01, 0.01]. }
    \label{fig:spod_modes}
\end{figure}

Quantitatively similar results are observed for the 3-D SPOD modes. Based on the eigenspectrum at $\beta/\beta_1 \approx 11$ (see figure~\ref{fig:spod_spectra}(c)), $\lambda_1$ is marginally separated from $\lambda_2$ for $St = 3.27$. Consequently, the leading mode at this frequency appears primarily in the SPSL region, with prominent structures observed in the USL region close to the SERN trailing edge, as shown in figure~\ref{fig:spod_modes}(b). The low-rank behavior is not observed at any other frequency for the 3-D case, and the wavepackets appear in multiple shear layers for Modes 1, 2, and 3.

In summary, the wavepackets are predominantly concentrated in a single shear layer when SPOD eigenvalues are well separated. The SPSL region is governed by dominant and higher frequencies ($St \geq 3.27$), whereas the low frequencies ($St \leq 2.5$) drive the dynamics in the USL and LSL regions. These results, obtained through data-driven SPOD, are further compared with the resolvent model in Section~\ref{sec:SPODvsResl}.

\subsection{Instabilities of the base flow} \label{sec:stability}
A study of instabilities is essential for selecting an appropriate discounted (temporal) window in the subsequent resolvent analysis and to ensure that the results are both meaningful and physically interpretable~\cite{sun2017}.
The stability analysis is conducted with the time-averaged mean flow at the center plane as the base state for a range of spanwise wavenumbers. The choice of mean flow as the base state in stability analysis has been discussed in previous studies~\cite{barkley2006, turton2015, beneddine2016, Sun:TCFD16}, and it usually provides a reliable prediction of the dominant temporal frequency in unsteady flows. The time-averaged flow is interpolated onto a mesh comprising approximately $0.2 \times 10^6$ grid points. The construction of the linear operator ${L}_{\overline{ q}; \beta}$ is carried out using an in-house finite-volume compressible flow solver~\cite{sun2017, bres2017}. Boundary conditions for perturbation density and velocity are set to zero at the inlet, outlet, and far-field boundaries, as well as at the nozzle walls, splitter plate, and aft-deck wall. Pressure is subjected to Neumann boundary conditions with zero normal gradients at all boundaries. Sponge layers are applied at the far-field and outlet boundaries to dampen outgoing waves and prevent reflections. As shown in Appendix~\ref{sec:Re_effect}, the Reynolds number does not significantly influence the dominant amplification mechanism of the linear operator. Thus, a Reynolds number of $Re = 1 \times 10^5$ is used to construct ${L}_{\overline{ q}; \beta}$. The resulting linear operator ${L}_{\overline{ q}; \beta}$, represented as a matrix, has an approximate size of $10^6 \times 10^6$. The frequency $\lambda_{Lr}$, the real component of the eigenvalue $\lambda_L$, is normalized as the Strouhal number $St = \lambda_{Lr} D_h / (2 \pi \overline{u}_j)$, while the growth/decay rate, the imaginary component of the eigenvalue $\lambda_L$, is normalized as $\lambda_{Li} D_h / \overline{u}_j$.

The eigenspectrum of the linear operator $L_{\overline{ q}; \beta}$ is shown in figure~\ref{fig:stability}(a). The black dashed line separates the stable region ($\lambda_{Li} D_h / \overline{u}_j<0$) from the unstable region ($\lambda_{Li} D_h / \overline{u}_j>0$) in the complex plane. The linear operator $L_{\overline{ q}; \beta}$ exhibits instability, with two unstable branches (mode-i and mode-ii branches shown by dashed-dotted lines in figure~\ref{fig:stability}(a)), observed at approximately $St \approx 0.5$ and 1.1 for 3-D instabilities, while 2-D modes remain stable. For the mode-i branch, its growth rate decreases as $\beta/\beta_1$ increases. When $\beta/\beta_1$ reaches 18, the leading eigenvalue shifts into the stable region. Conversely, the growth rate of mode-ii branch, with $St \approx 1.1$, increases as $\beta/\beta_1$ increases. The modal structures of the real component of $\hat{u}$ are shown in figure~\ref{fig:stability}(b) for the two unstable eigenvalues highlighted in figure~\ref{fig:stability}(a). The most unstable mode-i, characterized by $(St, \lambda_{Li} D_h / \overline{u}_j, \beta/\beta_1) = (0.45, 0.215, 12)$, displays a distinctive structure concentrated along the shear-layer region of the splitter plate. Similarly, the most unstable mode-ii, with $(St, \lambda_{Li} D_h / \overline{u}_j, \beta/\beta_1) = (1.18, 0.369, 24)$, is also concentrated in the shear layer of the splitter plate, but with a smaller wavelength due to its higher temporal frequency.
\begin{figure}[hbpt]
     \centering
        \includegraphics[width=0.35\textwidth]{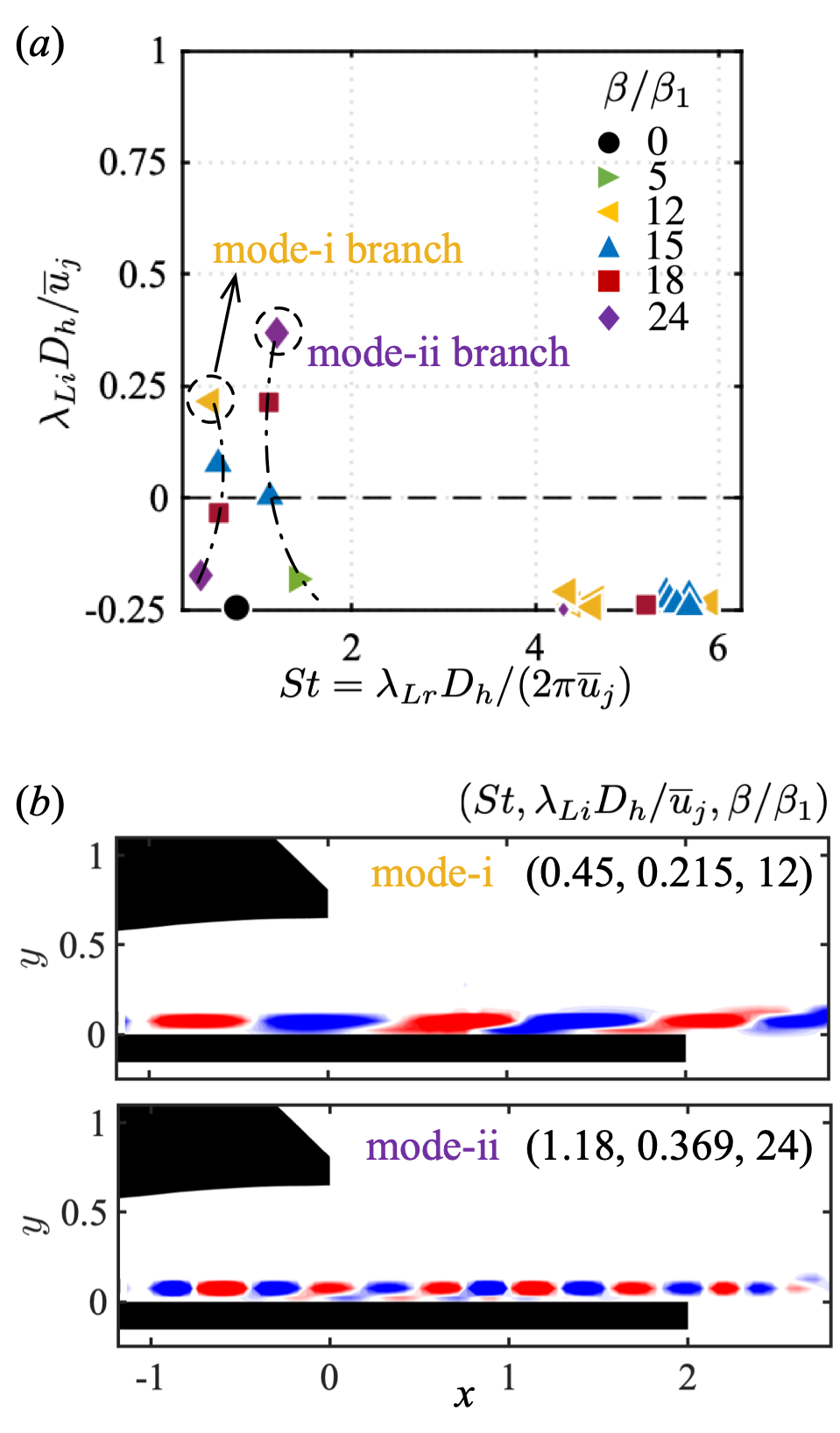}
    \caption{(a) Eigenvalues of $L_{\overline{ {q}}; \beta}$ for $0 \leq \beta/\beta_1 \leq 24$. (b) Modal structures (real component of $\hat{u}$) of two leading eigenmodes circled in the eigenspectrum plot. [blue, red] = [-0.001, 0.001]. }
    \label{fig:stability}
\end{figure}
The linear operator ${L}_{\overline{ q}; \beta}$ exhibits instability over a range of $\beta/\beta_1$, as evidenced by the presence of eigenvalues with positive growth rates.

\subsection{Resolvent analysis without any restrictions} \label{sec:resl_result}

The resolvent analysis is performed with the randomized algorithm~\cite{halko2011, ribeiro2020} to reduce the computational cost and memory requirements associated with the SVD. The gain distribution and modal structures obtained from both full SVD and randomized SVD demonstrate excellent agreement, as detailed in Appendix~\ref{sec:rsvd}. As discussed earlier, we adopt $Re = 1 \times 10^5$, since the amplification mechanism of the multi-stream jet remains unaffected by the Reynolds number (see Appendix~\ref{sec:Re_effect}). Consequently, the Reynolds number can be treated as a free parameter, similar to the approach taken by \cite{schmidt2018}.

\subsubsection{Effect of discounted parameter} \label{sec:dis_paramter}
We first investigate the effect of the discounted parameter ($\alpha$) on energy amplification for $\beta/\beta_1=0$. Figure~\ref{fig:gain_mode_disc}(a) presents the optimal gain for various values of the discounted parameter. The highest gain is observed at $St \approx 3.4$ with $\alpha D_h / \overline{u}_j = 0$. The resolvent gain arises from two contributions:  resonant mechanisms, when the input frequency ($\omega + i\alpha$) corresponds to an eigenvalue ($\lambda_L$) of the linear operator, and pseudoresonance associated with the non-normality of the eigenvectors of the non-normal operator~\cite{trefethen1991, schmid2007}. Since the stability analysis does not reveal any unstable eigenvalue at the dominant frequency $St=3.27$ for $\beta/\beta_1=0$,  we can say that the highest gain originates from the non-normality of the linear operator at frequency $St \approx 3.4$. 

To highlight key features, we plot the optimal gain distribution for three values of $\alpha D_h / \overline{u}_j = 0$, 1.57, and 3.14 in figure~\ref{fig:gain_mode_disc}(b). The leading gain decreases as the discounted parameter increases, primarily because the time window within which disturbances are allowed to grow becomes shorter with larger values of $\alpha D_h / \overline{u}_j$. Small peaks for $St \leq 1.5$ and $\alpha D_h / \overline{u}_j = 0$ are observed due to resonance with eigenvalues. As the discounted parameter increases, the complex line about which the gain is calculated moves away from the eigenvalues, resulting in a reduction of gain for $St \leq 1.5$. However, the overall gain distribution pattern near the dominant frequency remains unchanged.
\begin{figure} [hbpt]
    \centering
         \includegraphics[width=0.5\textwidth]{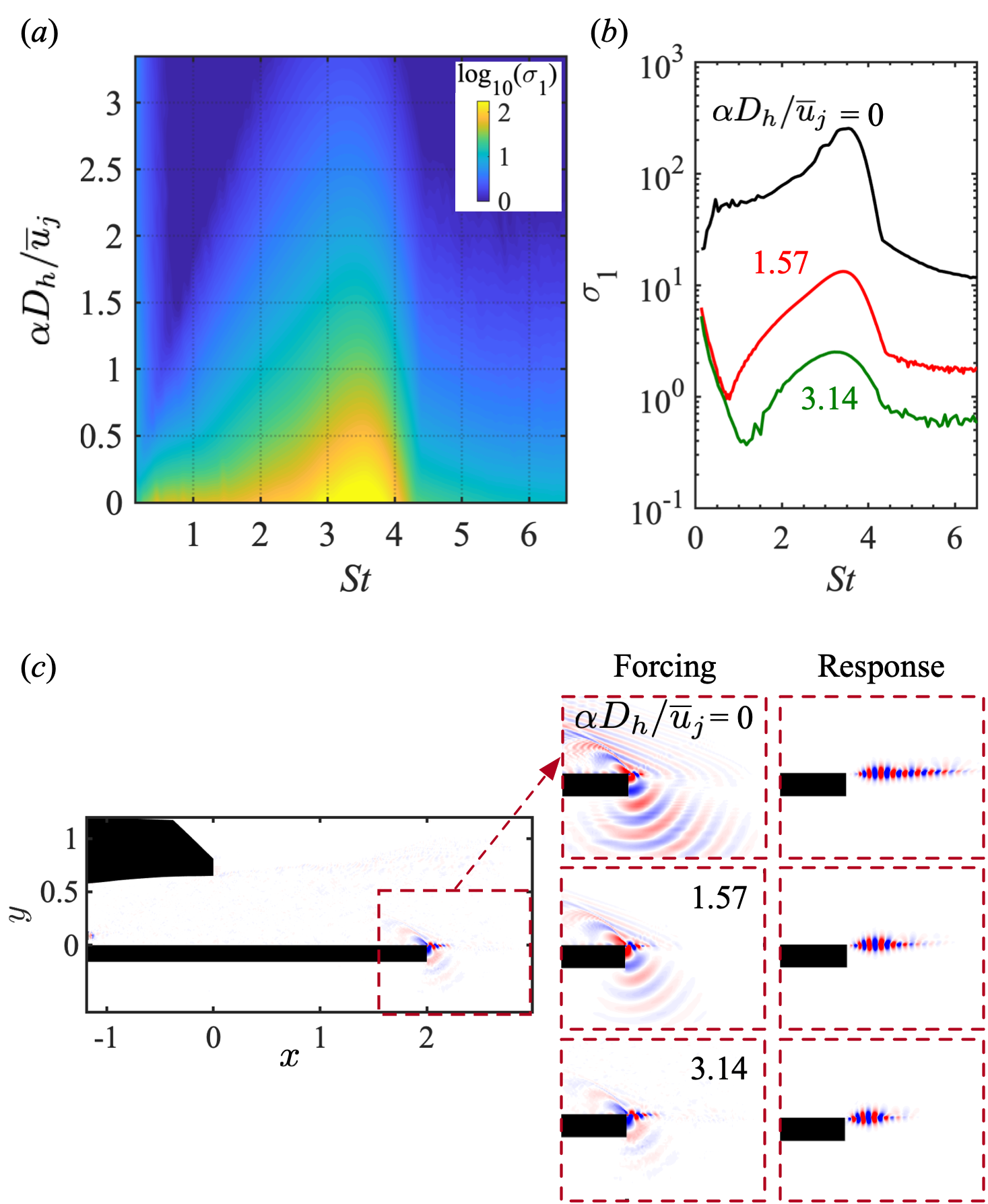}
    \caption{(a) Contour plots of optimal singular value ($\sigma_1$) for different $\alpha D_h/\overline{u}_j$ and $St$ with $\beta / \beta_1 = 0$. (b) The trend of $\sigma_1$ at $\alpha D_h/\overline{u}_j =$ 0, 1.57, and 3.14. (c) The optimal pair of forcing and response mode (real part of $\hat{v}$) are shown for $St = 3.5$, $\beta/\beta_1 = 0$ and $\alpha D_h/\overline{u}_j$ = 0, 1.57, and 3.14. Forcing \& response: [blue, red] = [-0.2,0.2].}
    \label{fig:gain_mode_disc}
\end{figure}

The spatial distributions of the modal structures are crucial in designing a practical flow control strategy, as they identify sensitive locations to introduce perturbations or sense the flow state. The optimal forcing and response modes for the representative values of $\alpha D_h/\overline{u}_j$ at $St = 3.5$ and $\beta/\beta_1 = 0$ are shown in figure~\ref{fig:gain_mode_disc}($c$). The optimal mode structures are located in the LSL region at the trailing edge of the aft deck for all $\alpha D_h/\overline{u}_j$. While the leading 2-D SPOD mode appears in the SPSL at the same frequency, as explained in Section~\ref{sec:SPODvsResl}, this discrepancy is due to the over-optimization of the resolvent operator.
The forcing and response modes contract toward the trailing edge of the aft deck as the discounted parameter increases, and the time window becomes shorter. With increasing $\alpha D_h/\overline{u}_j$, the forcing modes concentrate on the upper side of the trailing edge of the aft deck, pinpointing the sensitive region for input introduction.
Varying discounted parameters can also reveal different amplification mechanisms for a flow~\cite{garnaud2013, arratia2013, yeh2019}, although in the present work, the optimal forcing and response modes remain similar, indicating the same amplification mechanism across all discounted values. For the resolvent analysis, we choose $\alpha D_h/\overline{u}_j = 1.57$, as all unstable eigenvalues associated with each $\beta/\beta_1$ lie below this line, i.e., $\alpha D_h/\overline{u}_j > \text{max} (\lambda_{Li} D_h / \overline{u}_j)$, as shown in figure~\ref{fig:stability}(a).

\subsubsection{Amplification affected by spanwise wavenumbers} \label{sec:res_finite_time}

The optimal singular value ($\sigma_1$) of the resolvent operator for a parametric sweep through $St$ and $\beta/\beta_1$, with a fixed discounted parameter $\alpha D_h/\overline{u}_j = 1.57$, is shown in figure~\ref{fig:beta_gain}(a). Higher gains are observed at $St=3.4$ for all $\beta/\beta_1$. As discussed earlier, this is close to the dominant vortex shedding frequency $St = 3.27$ (corresponding to a dimensional frequency of approximately $33$ kHz) identified in LES~\cite{stack2018} and far-field acoustic spectra from experiments~\cite{berry2016}. Previous studies on isolated shear-layer flows similarly identified the vortex-shedding frequency as the optimal amplification frequency~\cite{yeh2020, doshi2022, thakor2024}. The value of $\sigma_1$ near the dominant frequency $St = 3.27$ decreases monotonically with increasing $\beta/\beta_1$. Figure~\ref{fig:beta_gain}(b) presents the first sub-optimal (second) singular value ($\sigma_2$). Within the range $3 \lesssim St \lesssim 4$ for $\beta/\beta_1 \leq 8$, $\sigma_2$ is approximately an order of magnitude lower than $\sigma_1$. In this parameter range, the rank-1 assumption in the resolvent analysis remains valid. However, outside this range, the resolvent modes are no longer rank-1. Consequently, sub-optimal modes need attention to better understand the underlying physics or to develop reduced-order models that accurately describe the original flow.
\begin{figure}[hbpt]
    \centering
         \includegraphics[width=0.38\textwidth]{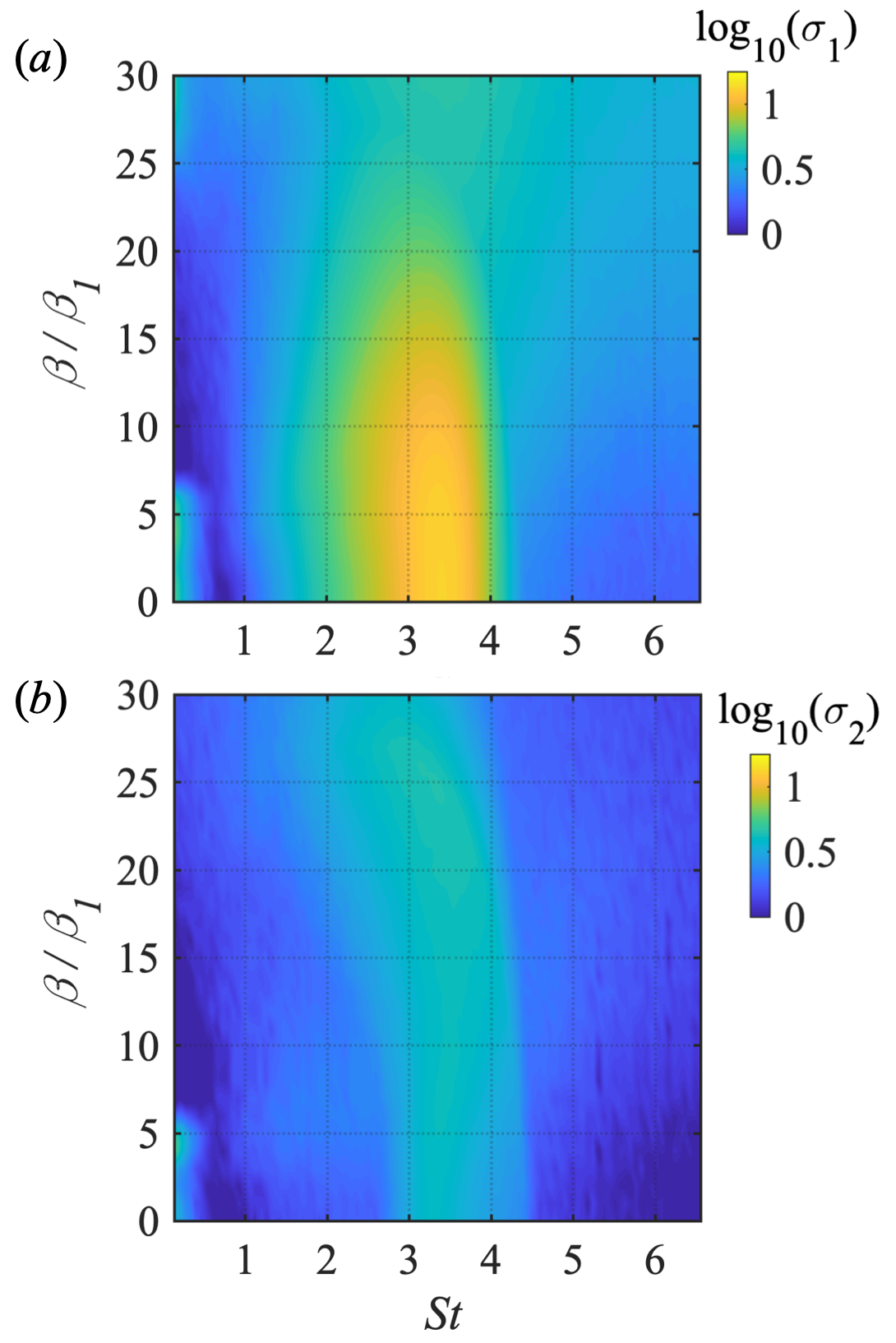}
    \caption{Contour plots of (a) the optimal singular value $\sigma_1$, and (b) the sub-optimal singular value $\sigma_2$ of the resolvent operator with $\alpha D_h/\overline{u}_j = 1.57$.}
    \label{fig:beta_gain}
\end{figure}

Additionally, figure~\ref{fig:beta_gain_line} shows the first three singular values for representative spanwise wavenumbers. Notably, the singular values exhibit a \textit{shift} or crossover in their path at specific frequencies. The locations where this occurs are marked with red ($\sigma_1 \leftrightarrow \sigma_2$) and blue ($\sigma_2 \leftrightarrow  \sigma_3$) circles. This phenomenon also appears in input-output analysis, where non-identity input/output matrices are introduced. A detailed discussion on the relationship between this singular value \textit{shift} and the corresponding modal structures is provided in Section~\ref{sec:io_state}.
\begin{figure}[hbpt]
     \centering
         \includegraphics[width=1\textwidth]{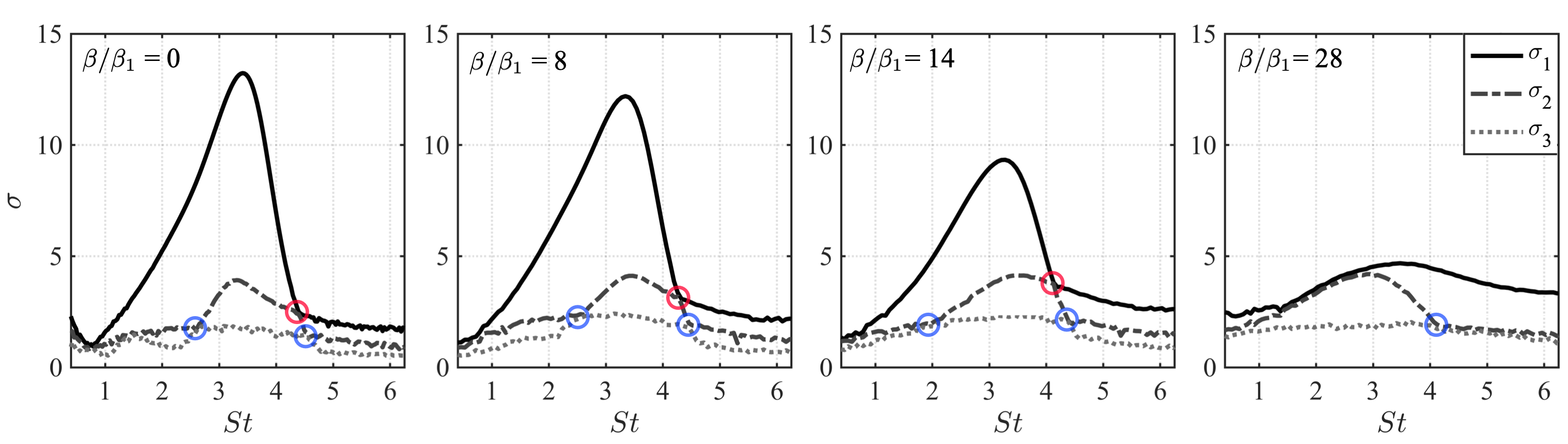}
    \caption{ The optimal ($\sigma_1$) and sub-optimal ($\sigma_2$ and $\sigma_3$) singular values at the representative spanwise wavenumber. The red and blue circles indicate the location of \textit{shift} between $\sigma_1 \leftrightarrow \sigma_2$ and $\sigma_2 \leftrightarrow  \sigma_3$, respectively.}
    \label{fig:beta_gain_line}
\end{figure}

The first three forcing (as green-yellow line contours) and response (as blue-red flooded contours) modes (real part of $v$-velocity) are depicted in figure~\ref{fig:mode_beta_st} for representative pairs of ($St$, $\beta/\beta_1$).
The optimal 2-D modes ($\beta/\beta_1=0$) are located in the LSL for $St=2.1$ and $St=3.27$, as shown in figure~\ref{fig:mode_beta_st}(a). Since this flow configuration includes multiple shear layers, the dynamics of the other shear layers are captured in the sub-optimal modes. However, the input frequency determines the optimal response among three shear layers for a specific wavenumber. For instance, at a higher frequency ($St=5.76$), the optimal forcing-response pair emerges in the SPSL, indicating a transition in the dominant forcing-response location from the LSL to the SPSL between frequencies $St=3.27$ and 5.76. A similar transition is observed in the sub-optimal mode~2, which shifts from the USL to the SPSL between $St=2.1$ and 3.27. For sub-optimal mode~3, the dominant structures appear in the USL at $St=3.27$. In contrast, the modes for $St=2.1$ and $St=5.76$ exhibit noisy structures. 

We also examine the effect of three-dimensionality on the modal structure in figure~\ref{fig:mode_beta_st}(b). For 3-D modes, the spatial structures of the forcing and response become more compact in the transverse direction. At very high spanwise wavenumber ($\beta/\beta_1 = 28$), the optimal forcing-response pair consistently appears in the SPSL across all frequencies, as shown in figure~\ref{fig:mode_beta_st}(b). In contrast, the sub-optimal modes highlight other shear-layer regions.
These observations indicate that the optimal location for introducing perturbations, and thereby achieving the maximum flow response, depends on the specific combination of frequency and spanwise wavenumber.
\begin{figure} [hbpt]
    \centering
         \includegraphics[width=1\textwidth]{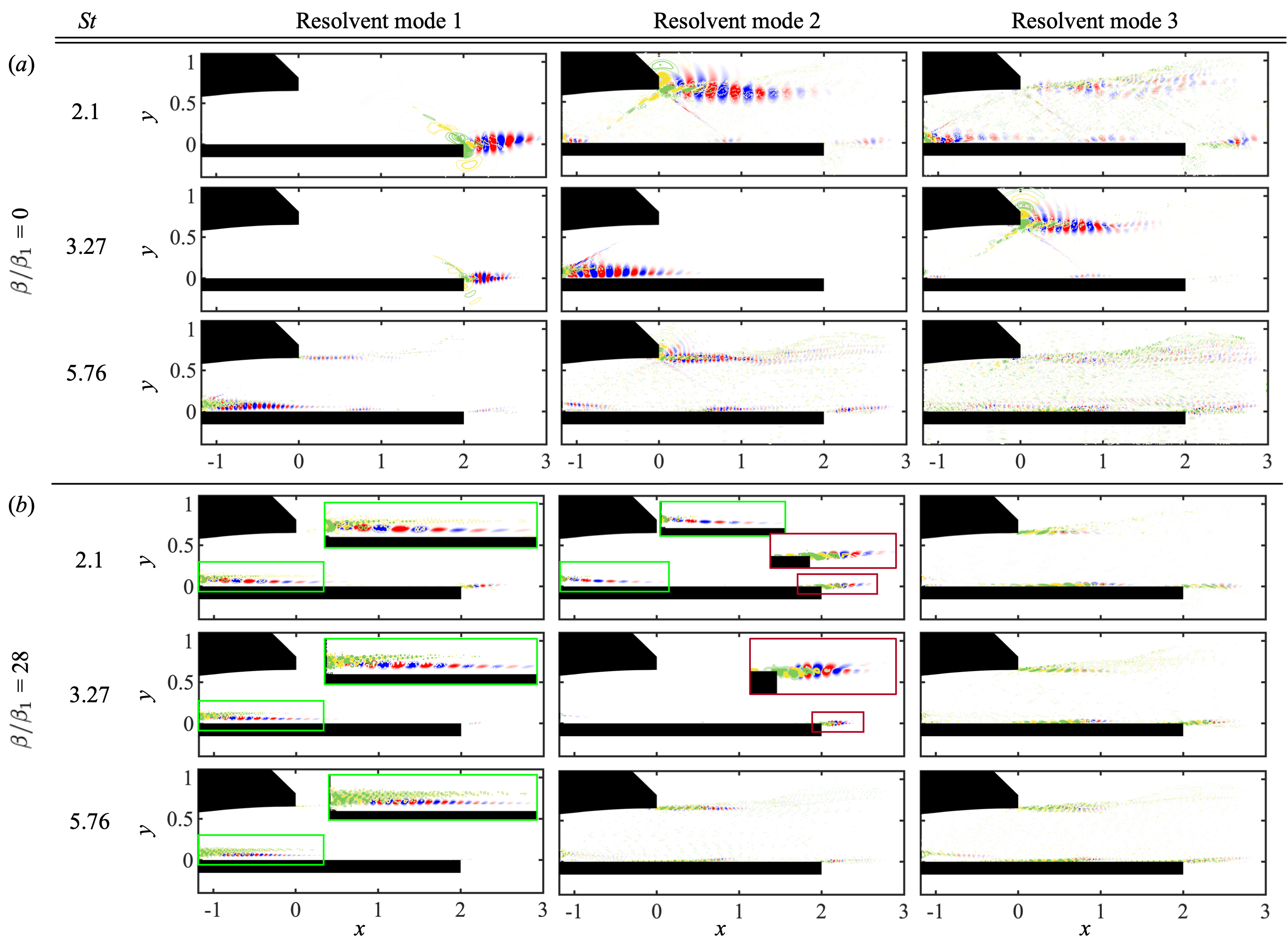}
    \caption{The pair of first three forcing and response modes (real component of $\hat{v}$) for the representative frequencies at $\beta/\beta_1 = $ (a) 0 and (b) 28. The zoom window near the lower shear layer (red color) and the splitter plate shear layer (green color) are shown for better visualization. Forcing: [green, yellow] = [-0.2, 0.2]. Response: [blue, red] = [-0.2, 0.2].}
    \label{fig:mode_beta_st}
\end{figure}

To quantify the optimal spatial amplification mechanism based on these observations, we consider three spatial windows corresponding to the SPSL, LSL, and USL to isolate the modal structures within their respective regions, as shown in figure~\ref{fig:dominant_region}(a). The momentum mixing is evaluated by performing a spatial integral of the corresponding Reynolds stresses over the region of interest. The Reynolds stresses are derived from the resolvent response modes~\cite{nakashima2017}. The spatial integration for modal stress is expressed as,
\begin{equation}
M(St, \beta/\beta_1) = \int_S [\sigma_1^2 ( \hat{R}_x^2 + \hat{R}_y^2 + \hat{R}_z^2)^{1/2}] dS.
\label{eq:momentum_mixing}
\end{equation}
Here, $S$ represents the respective windows, and the modal Reynolds stresses are defined as $ \hat{R}_x = \text{real} (\hat{v}^* \hat{w})$, $ \hat{R}_y = \text{real} (\hat{w}^* \hat{u})$, and $ \hat{R}_z = \text{real} (\hat{u}^* \hat{v})$. Figure~\ref{fig:dominant_region}(b-d) illustrates the normalized momentum mixing, $\hat{M} = M_i/\sum M_i$, ($i =$ LSL, SPSL, USL), computed using Eq.~\ref{eq:momentum_mixing} within the respective regions. The dark red color highlights the dominant spatial location based on the optimal response mode. The response mode is concentrated in the LSL for the frequency-wavenumber range $1 < St < 4.5$ and $\beta/\beta_1 < 28$, as depicted in figure~\ref{fig:dominant_region}(b). In contrast, for higher frequencies, $St \geq 4.5$, the optimal receptive region shifts to the SPSL. At very high spatial wavenumbers, $\beta/\beta_1 \geq 28$, the optimal response mode is consistently located in the SPSL for all temporal frequencies (figure~\ref{fig:dominant_region}(c)). The USL is predominantly responsive at low frequencies ($St \leq 0.5$) and for low-to-mid wavenumber pairs ($\beta/\beta_1 < 20$), as shown in figure~\ref{fig:dominant_region}(d). This analysis clearly demonstrates that the optimal receptive region in this rectangular jet flow with multiple shear layers is contingent on the specific frequency and spanwise wavenumber pair.
\begin{figure}[hbpt]
    \centering
         \includegraphics[width=1\textwidth]{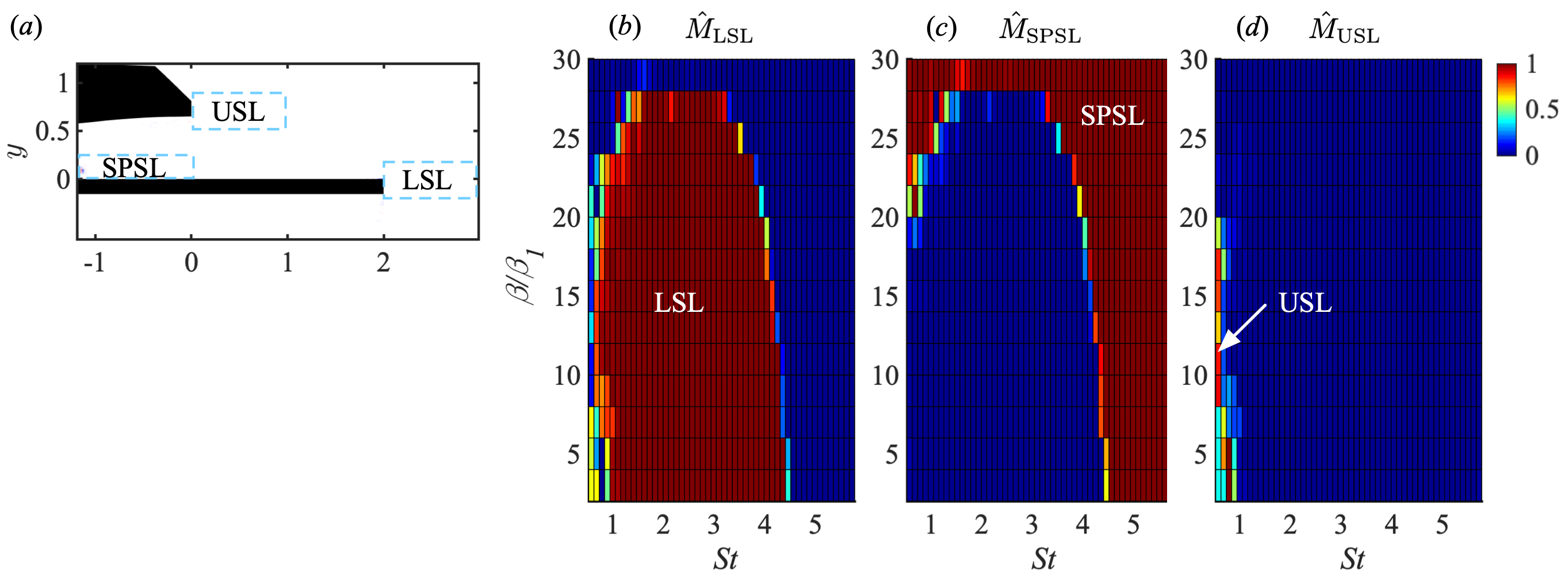}
    \caption{ (a) Three spatial windows in which modal Reynolds stresses are calculated. Normalized momentum mixing ($\hat{M}$) in (b) lower shear layer (LSL), (c) splitter plate shear layer (SPSL), and (d) upper shear layer (USL).}
    \label{fig:dominant_region}
\end{figure}

\subsection{Comparison of the resolvent model to SPOD method} \label{sec:SPODvsResl}

We compare the results from the resolvent model with those from SPOD, referring to the known connections between them~\cite{towne2018}. Theoretically, the resolvent and SPOD yield identical modal structures for white noise forcing i.e.,  uncorrelated in space and time. However, a one-to-one correspondence between resolvent and SPOD is not expected, as the forcing is typically not white in a highly turbulent flow. By comparing the leading mechanisms of these two methods, we expect the higher-gain mechanisms identified by the resolvent model to be compatible with the high-energy SPOD modes.

Figure~\ref{fig:spod_resl_spectra} shows the first three leading SPOD energies ($\lambda$, as a percentage of total energy) and resolvent amplification energies ($\sigma^2$) for $\beta/\beta_1 = 0$ and 11. The optimal resolvent spectral peaks lie near the dominant frequency $St = 3.27$; this behavior is similar that of the leading SPOD spectra for both the 2-D and 3-D cases. Interestingly, the first sub-optimal resolvent energy ($\sigma_2^2$) also exhibits a peak at the dominant frequency $St = 3.27$ in both 2-D and 3-D cases, indicating that two directions corresponding to $\sigma_1$ and $\sigma_2$ are preferentially amplified by the linear operator. Meanwhile, the first SPOD eigenvalue captures a significant portion of the energy. Another notable difference is that the resolvent peaks are more broadband compared to the sharper SPOD peaks at $St = 3.27$.
\begin{figure}[hbpt]
     \centering
         \includegraphics[width=1\textwidth]{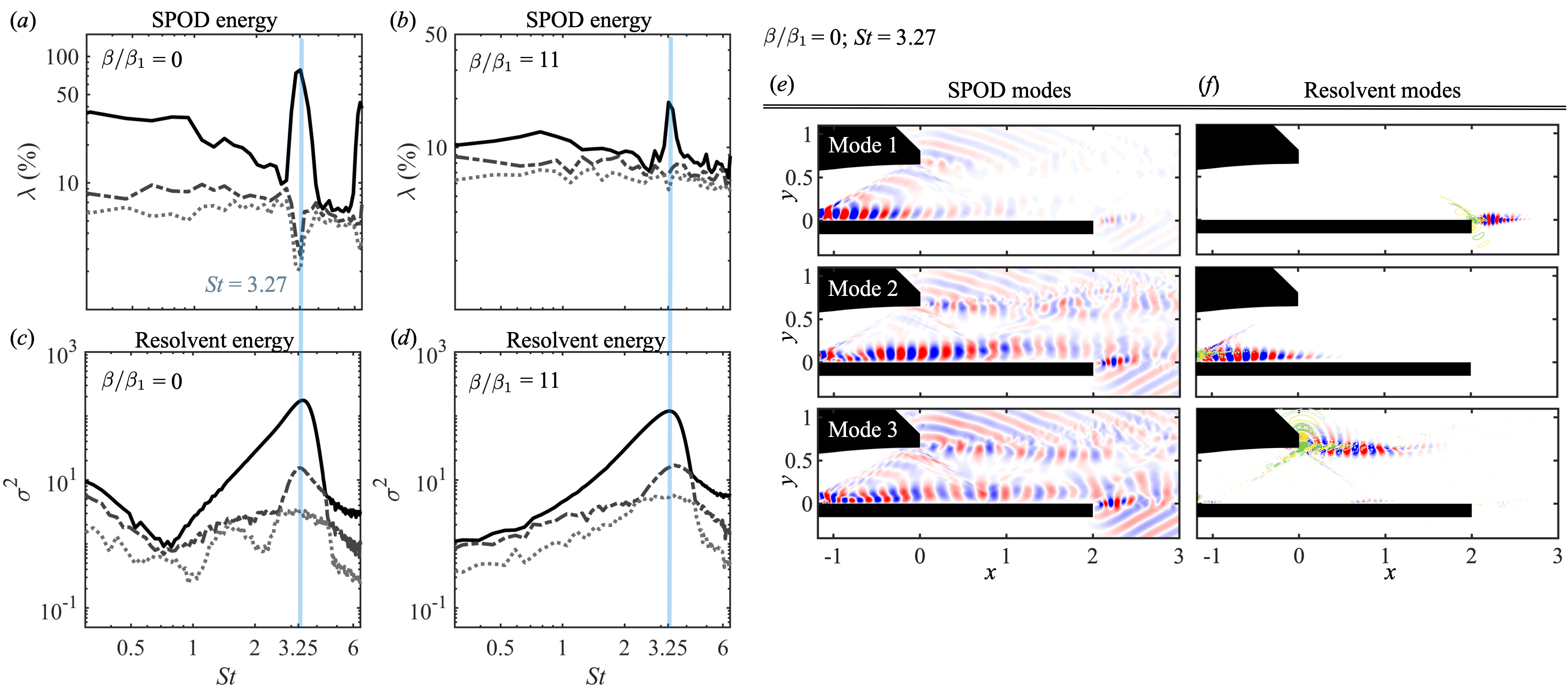}
    \caption{The first three SPOD energy spectra at $\beta/\beta_1 = $ (a) 0, (b) 11; and the first three resolvent gain at $\beta/\beta_1 = $ (c) 0, (d) 11. $\lambda_1, \sigma_1:$ Solid, $\lambda_2, \sigma_2:$ Dash, and $\lambda_3, \sigma_3:$ Dot lines. The comparison of (e) the first three SPOD and (f) resolvent modes at $St=3.27$ and $\beta/\beta_1=0$. The colormaps for SPOD and resolvent are consistent with those in the previous section. }
    \label{fig:spod_resl_spectra}
\end{figure}

Next, the first three SPOD modes are plotted against the resolvent modes at the dominant frequency $St = 3.27$ and $\beta/\beta_1 = 0$ in figure~\ref{fig:spod_resl_spectra}(e) and \ref{fig:spod_resl_spectra}(f), respectively. The wavepackets are observed in the SPSL region for the leading SPOD mode, whereas the optimal resolvent mode predicts wavepackets in the LSL region, with the first sub-optimal mode (Mode~2) appearing in the SPSL region. This discrepancy arises from a modeling assumption of the linear resolvent operator as described by~\cite{mckeon2010}. Since the singular values of the resolvent operator rank are based on the amplitude of the response mode per unit forcing, a selection process occurs at each frequency-wavenumber pair. From broadband forcing, the forcing component in the direction corresponding to the first singular value is dominantly amplified. Consequently, the linear operator maps this forcing to produce the optimal response mode. It is, however, not necessary for the optimal response to fully describe the current flow at the given frequency-wavenumber pair, as the forcing is arbitrary. In our case, where multiple shear layers are present, the linear mechanism over-optimizes the forcing-response, predicting optimal wavepackets in the LSL at $St = 3.27$. As discussed earlier, the resolvent operator also amplifies $\sigma_2$, where the response appears in the SPSL (resolvent Mode 2), consistent with SPOD Mode 1. Despite this over-optimization, the wavelength of the modes predicted by the resolvent analysis aligns completely with the convective wavepackets in the SPSL region. In the following section, we will demonstrate that the optimal resolvent mode becomes more aligned with the leading SPOD mode when the optimization is performed over a subspace of the resolvent operator. To fully understand the discrepancies in capturing leading modes from the two methods, various resolvent formulations with forcing models (for example, see~\cite{pickering2021, skene2022, kamal2023, bugeat2024}) may potentially improve the linear prediction in such complex flows.

\subsection{Componentwise input-output analysis} \label{IO}

This section examines various combinations of input and output configurations based on the state variables (i.e., velocity components and temperature) of interest, motivated by the challenges and constraints in active flow control design. In practical scenarios, actuators are typically placed on solid surfaces with specific forcing inputs~\cite{glezer2002, cattafesta2011}. For all results in this section, the discounted parameter is set to $\alpha D_h/\overline{u}_j = 0.78$, which is greater than the most unstable eigenvalue, i.e., $(\alpha D_h/\overline{u}_j) > \text{max} (\lambda_{Li} D_h / \overline{u}_j)$.

\subsubsection{Effect of an input/output state restriction} \label{sec:io_state}

The input-output analysis is performed by manipulating the input ($B$) and output ($C$) matrices to restrict the state variable, enabling the examination of specific dynamics among particular variables. The optimal singular value ($\sigma_1$) is presented in figure~\ref{fig:io_gain} for all the cases considered. The label $\Tilde{  {H}}_{u \rightarrow v}$ represents the input provided in the streamwise velocity ($u$), with the output observed in the transverse velocity ($v$). This notation is used consistently across all other cases. A red dot in each plot indicates the maximum gain for each individual case.
\begin{figure}[hbpt]
     \centering
         \includegraphics[width=1\textwidth]{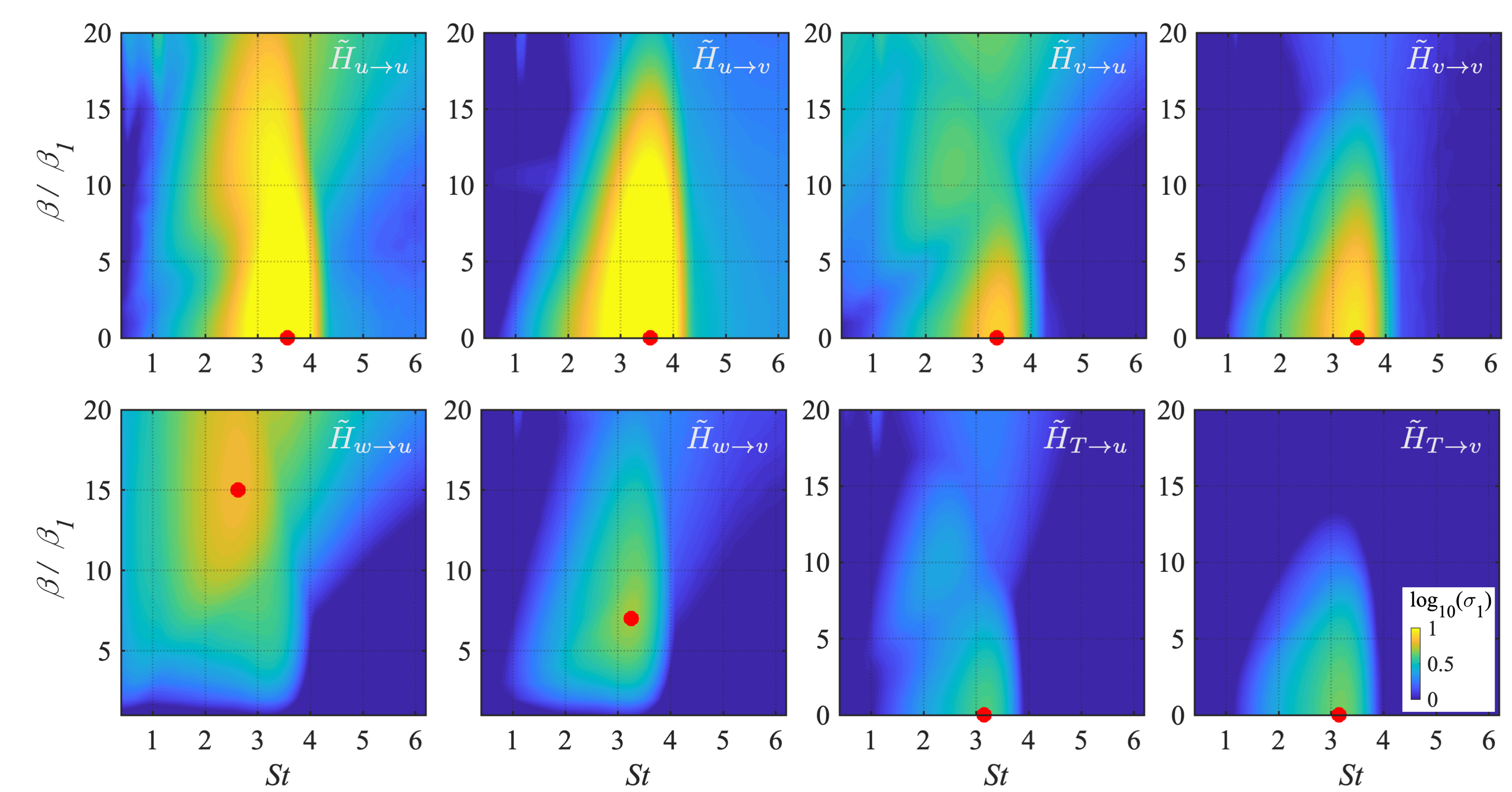}
    \caption{Contour of leading gain as a function of $St$ and $\beta/\beta_1$ for componentwise input-output analysis. Label of $\Tilde{  {H}}_{u \rightarrow v}$ presents the case with only streamwise velocity ($u$) as input and only transverse velocity ($v$) as output. This notation is used for all the other cases. The red dot in each plot indicates the maximum gain.}
    \label{fig:io_gain}
\end{figure}


The 2-D mechanism ($\beta/\beta_1=0$) produces the largest energy amplification near the dominant frequency for individual cases when input is provided at the $u$-, $v$-velocities, and $T$. Comparing across all cases, the maximum gain is observed for $\Tilde{  {H}}_{u \rightarrow v}$ at $St \approx 3.5$ and $\beta/\beta_1=0$, indicating that the 2-D input in the $u$-velocity at this temporal frequency generates the maximum flow response among all the cases considered. We observe that the gain monotonically decreases with an increase in wavenumber for cases where the output is restricted to the $v$-velocity. Conversely, the gain decreases and becomes more broadband for higher wavenumbers in all $u$-velocity output cases. The $w$-velocity input cases induce the highest relative amplification for non-zero wavenumber. For instance, the highest gain is observed at $(St, \beta/\beta_1) = (2.5, 15)$ for the $\Tilde{  {H}}_{w \rightarrow u}$ case. Moreover, the amplification observed from temperature ($T$) input to velocity outputs is significantly lower than that for velocity inputs. Therefore, component-wise analysis of energy amplification for various inputs and outputs reveals that velocity-based inputs are likely to be more effective than thermal-based inputs in modifying the flow. Specifically, the input of streamwise velocity ($u$) exhibits the highest amplification among all the cases.

As the rank-1 assumption is not always valid in such a complex flow with multiple shear layers, we examine the optimal and sub-optimal energy amplification (gain) distributions with different spanwise wavenumbers, focusing on the case $\Tilde{  {H}}_{v \rightarrow u}$. Figure~\ref{fig:vTOu_gain}(a) depicts the optimal (first singular value $\sigma_1$: solid line) and first sub-optimal (second singular value $\sigma_2$: dotted line) gains for representative values of $\beta/\beta_1 \leq 18$. The optimal gain is significantly higher than the first sub-optimal gain for $\beta/\beta_1 = 0$ around the dominant frequency $St = 3.27$, indicating that the 2-D mechanism is predominant in the flow for the input-output case as well. As the spanwise wavenumber increases, the difference between the first and second singular values diminishes, rendering the rank-1 assumption no longer valid. The gain distribution shifts to a more broadband type for higher $\beta/\beta_1$ values around the dominant frequency. A sharp peak is observed at $St \approx 1.1$ for $\beta/\beta_1 = 18$ due to resonance, as the eigenvalue exists at the same frequency, as shown in the stability spectrum (see figure~\ref{fig:stability}(a)).
\begin{figure}[hbpt]
     \centering
         \includegraphics[width=1\textwidth]{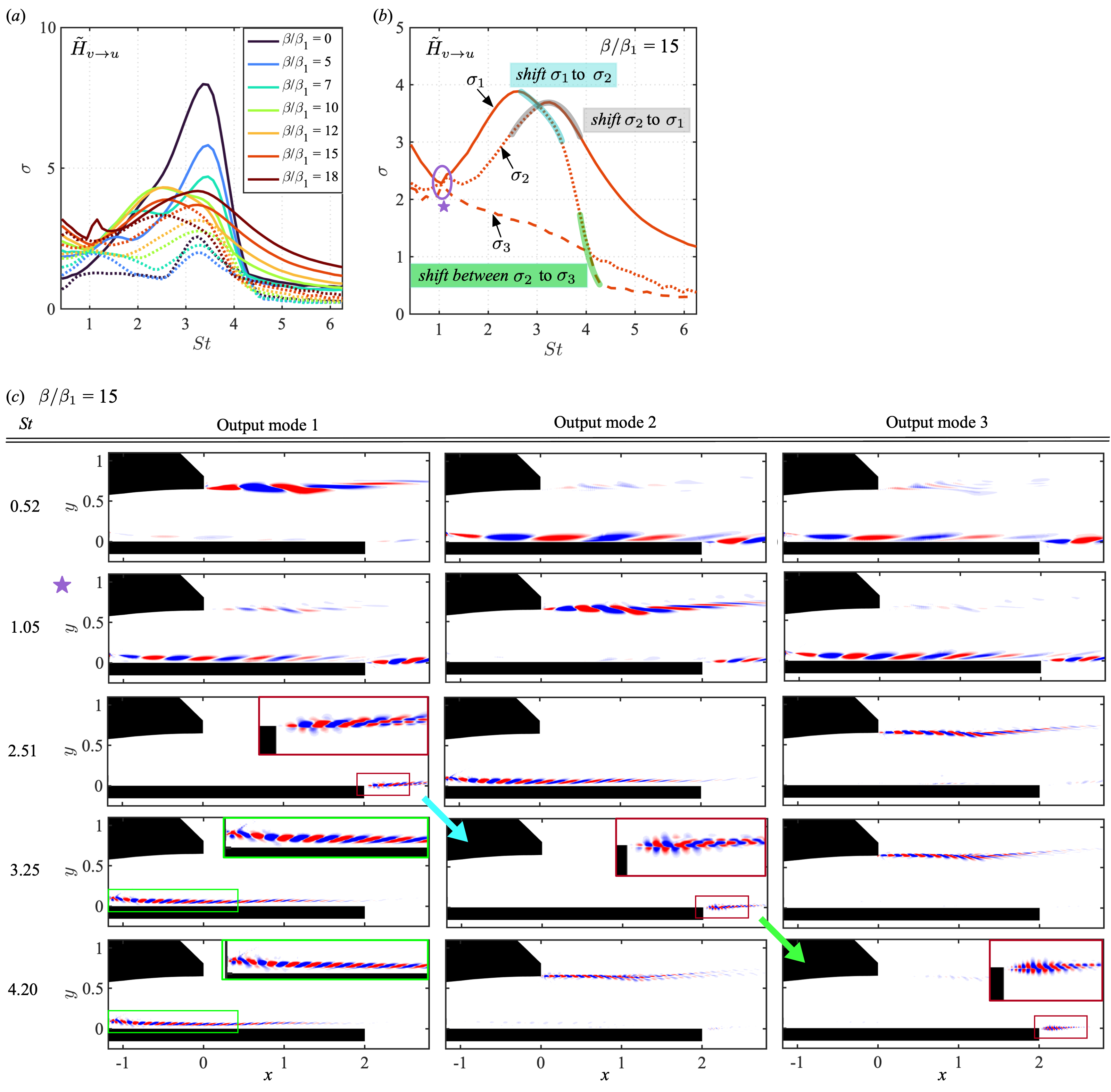}
    \caption{(a) The optimal (first singular value, $\sigma_1$) and suboptimal (second singular value, $\sigma_2$) are shown for $\Tilde{  {H}}_{v \rightarrow u}$ at $0 \leq \beta/\beta_1 \leq 18$. (b) The optimal ($\sigma_1$) and two suboptimal ($\sigma_2$ and $\sigma_3$) singular values for $\beta/\beta_1 = 15$ and $\Tilde{  {H}}_{v \rightarrow u}$. $\sigma_1:$ Solid, $\sigma_2:$ Dot, and $\sigma_3:$ Dash lines. The star sign denotes an overlap of the leading three singular values. (c) First three $u-$velocity (real component) output modes from $\Tilde{  {H}}_{v \rightarrow u}$ with $\beta/\beta_1 = 15$ at representative frequencies. The green and red zoom windows are shown in the top right corner for better visualization in the SPSL and LSL regions. [blue, red] = [-0.1, 0.1].}
    \label{fig:vTOu_gain}
\end{figure}

We further analyze the optimal and first two sub-optimal gain distributions, as the gain curve for the case with $\beta/\beta_1 = 15$ exhibits two local peaks, as shown in figure~\ref{fig:vTOu_gain}(b). The two peaks are observed in the optimal gain ($\sigma_1$) at approximately $St = 2.5$ and 3.2, and the optimal and first sub-optimal gains are identical at a frequency of $St \approx 3.1$. If we consider a smooth gain curve, $\sigma_1$ appears to follow a path shown in light blue, while $\sigma_2$ follows a path shown in gray. Thus, we can define a \textit{shift} that occurs between these two gain curves around $St \approx 3.1$. A similar \textit{shift} is observed for $\sigma_2$ and $\sigma_3$ at $St \approx 4.1$. Around $St \approx 1.1$, the optimal and first two sub-optimal gains are nearly identical, with a \textit{shift} occurring among all three gain curves. The subsequent sub-optimal gains ($\sigma_4, \sigma_5, \dots$) are clearly separated from the first three singular values, which are not shown here.

Figure~\ref{fig:vTOu_gain}(c) shows the first three output modes (real component of $u-$velocity) corresponding to $\sigma_1$, $\sigma_2$, and $\sigma_3$, respectively, at representative frequencies. The optimal output mode-1 highlights the shear-layer response in different regions depending on the input frequencies. For instance, output mode-1 concentrates in the USL for a low frequency $St = 0.52$, whereas the LSL is more responsive at $St = 2.51$. The modal structures are located in the SPSL for $St \geq 3.25$. This observation also aligns well with the previous discussion on the dominant amplification region in the flow field (Section~\ref{sec:res_finite_time}), although the frequencies differ.

Next, we will establish a relation between the gain \textit{shift} phenomenon and the corresponding change in the output modal structures. In the frequency range $1.1 < St < 3$, the first three singular values are clearly separated from each other (figure~\ref{fig:vTOu_gain}(b)). In this frequency range, output mode-1, mode-2, and mode-3 are located in the LSL, SPSL, and USL, corresponding to $\sigma_1$, $\sigma_2$, and $\sigma_3$, respectively, and each mode is confined to a single shear layer. This relationship is reflected in figure~\ref{fig:vTOu_gain}(c) for $St = 2.51$.
As described above, a \textit{shift} occurs between $\sigma_1$ and $\sigma_2$ at $St \approx 3.1$, and the optimal responsive region switches from the LSL to the SPSL. After this \textit{shift}, mode-1 and mode-2 switch shear layers (indicated by the blue arrow) and appear in the SPSL and LSL at $St = 3.25$, respectively. Similarly, a \textit{shift} and a mode location switch are also observed between $\sigma_2$ and $\sigma_3$ at $St \approx 4.1$. For instance, mode-2 and mode-3 are located in the LSL and USL at $St = 3.25$, and they switch their locations at $St = 4.2$ in the USL and LSL, respectively (indicated by the green arrow). We also examine scenarios where two or more singular values are nearly identical, resulting in a gain curve \textit{shift}. At $St = 1.05$, a \textit{shift} occurs among the first three singular values, as indicated by the star in figure~\ref{fig:vTOu_gain}(b). The optimal output mode-1 is observed in all three major shear layers -- SPSL, LSL, and USL -- at $St = 1.05$ (figure~\ref{fig:vTOu_gain}(c)). The sub-optimal modes, mode-2 and mode-3, are also located in multiple shear-layer regions for this particular frequency.

From this analysis, we find that the three shear-layer mechanisms located in different regions of the jet flow are decoupled from each other, with each exhibiting a smooth gain curve as a function of frequency. The dominance among the three shear-layer mechanisms is highly dependent on the forcing frequency and spanwise wavenumber. Active flow control strategies should be tailored according to the control objectives, taking into account the input location and the temporal-spatial frequencies, to effectively modulate different shear layer regions. 

\subsubsection{Effect of an input spatial restriction} \label{sec:io_space}

We further introduce spatial restrictions in the input matrix $B$, confining it to a square box with dimensions corresponding to the thickness of the splitter plate at the SPTE (see figure~\ref{fig:uTOu_space_gain}(a)). The selection of the input location is motivated by our recent finding for an isolated SPSL, where we identified the SPTE as the most receptive region to introduce perturbation~\cite{thakor2024}. Based on the earlier discussion, the forcing is imposed only in the streamwise velocity, for which maximum amplification is observed. The output is constrained in the streamwise velocity, but without any spatial restrictions.

The gain distribution with and without spatial constraints at the SPTE for $\Tilde{  {H}}_{u \rightarrow u}$ is shown in figures~\ref{fig:uTOu_space_gain}(a) and (b), respectively. The dominant frequency $St \approx 3.3$ yields the highest gain for all wavenumbers when forcing is restricted to the SPTE, as shown in figure~\ref{fig:uTOu_space_gain}(a). The resonant mechanism contributes to the two sharp peaks at $St \approx 1.1$ and $St \approx 0.5$. The ratio $\sigma_1/\sigma_2 \approx \mathcal{O}(10^1)$ holds for all frequencies, implying a strong rank-1 behavior. Hence, optimal mode-1 is likely to be sufficient to represent the physical mechanism in the flow. We also note that a \textit{shift} between $\sigma_1$ and $\sigma_2$ is not observed as it is in the case without spatial constraints (see figure~\ref{fig:uTOu_space_gain}(b)). The gain oscillates near $St=3.3$ based on the wavenumber when the input is constrained at the SPTE. For instance, $\sigma_1$ decreases as the wavenumber increases in the range $\beta/\beta_1 < 5$, and vice versa for $\beta/\beta_1 > 5$. Conversely, a monotonic decay is observed in the case without any spatial restrictions. For completeness, we also plot the optimal $u-$velocity output mode for $\beta/\beta_1 = 0$ and $\beta/\beta_1 = 18$ for the SPTE input in figure~\ref{fig:uTOu_space_gain}(c). As anticipated, the SPSL exhibits a strong response across all input temporal frequencies and wavenumbers. Focusing on key highlights, the 2-D mode resembles the pattern of the KH instabilities of the base flow in the SPSL region at $St=3.25$. Meanwhile, the 3-D response convects further in the SPSL than the 2-D response. The current analysis demonstrates that spatially restricting input to a tailored practical flow control strategy can identify different optimal parameters that can induce maximum response.
\begin{figure}[hbpt]
     \centering
         \includegraphics[width=1\textwidth]{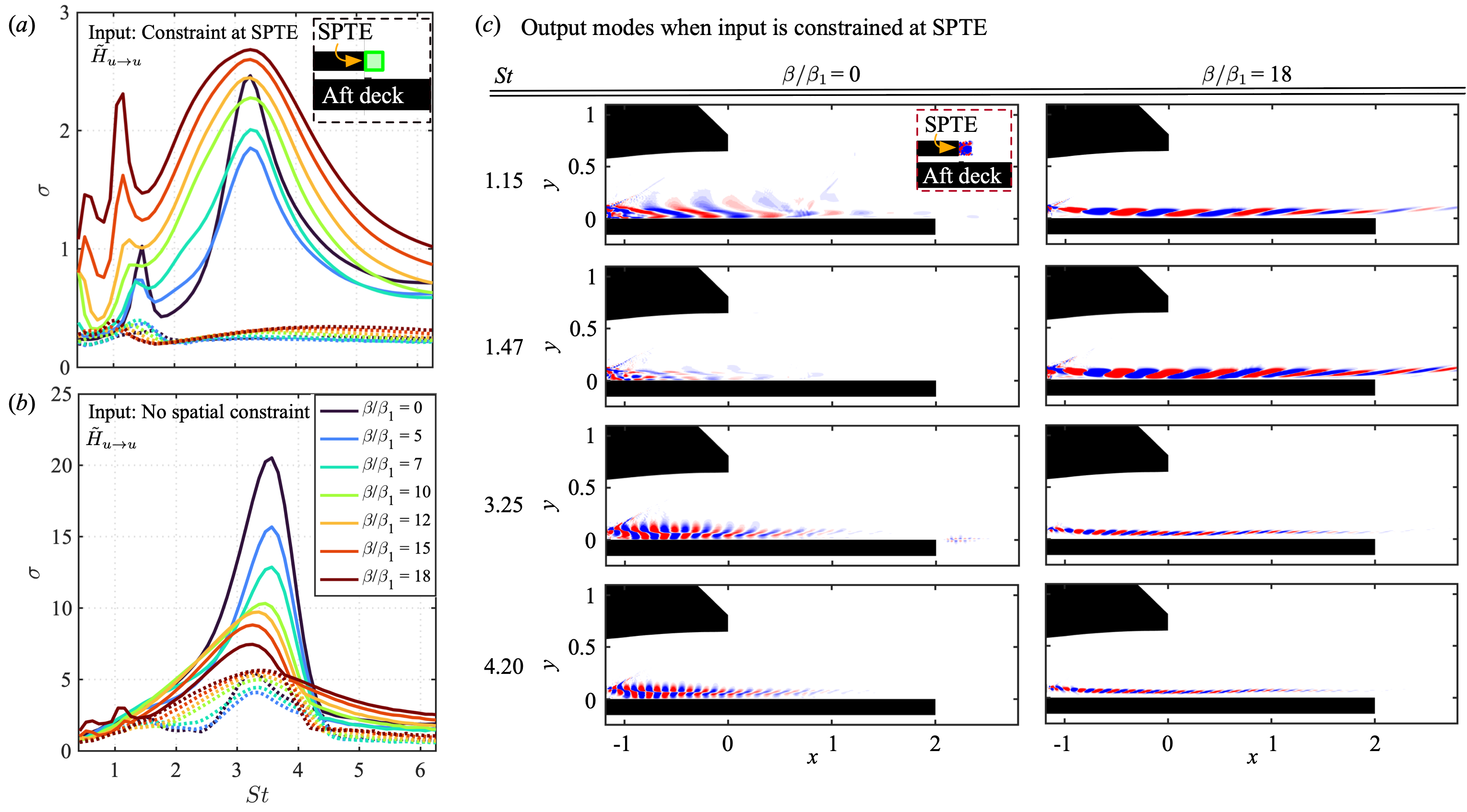}
    \caption{The optimal $\sigma_1$ and sub-optimal $\sigma_2$ singular values are shown for $\Tilde{  {H}}_{u \rightarrow u}$ at $0 \leq \beta/\beta_1 \leq 18$. (a) With spatial constraint, which is the square box (shown in green box) size of the splitter plate thickness at the SPTE, (b) Without any spatial constraint in input. $\sigma_1:$ Solid, $\sigma_2:$ Dot lines.}
    \label{fig:uTOu_space_gain}
\end{figure}


\section{Conclusion} \label{sec:conclusion}

We leverage linear operator-based methods to study the perturbation dynamics in a multi-stream rectangular supersonic jet flow. Previous numerical and experimental studies have reported high-frequency ($St=3.27$) loading on the aft-deck plate and generation of a far-field acoustic tone for this configuration. Spectral proper orthogonal decomposition (SPOD) reveals that the high-frequency phenomenon at $St=3.27$ is associated with 2-D Kelvin-Helmholtz (KH) vortex shedding in the splitter plate shear (SPSL) region, which arises from the mixing of the core and deck (bypass) streams. The coherent structures at lower frequencies ($St \leq 2.5$) are located in the upper- and lower-shear layer (USL and LSL) regions. SPOD further demonstrates that the 2-D mechanism dominates over the 3-D mechanism, with modal structures concentrated in a single shear layer when the SPOD energy spectrum is well separated from subsequent energy spectra.

Linear stability analysis identifies two unstable eigenvalue branches for non-zero wavenumbers. Subsequent discounted resolvent analysis examines the impact of the discounted parameter and spanwise wavenumber on energy amplification. Notably, a higher gain is observed at $St = 3.4$, close to the dominant vortex-shedding frequency, for all discounted parameters ($\alpha$), attributed to the non-normality of the linear operator. The optimal gain for various spanwise wavenumbers peaks near the dominant vortex-shedding frequency and diminishes rapidly at higher wavenumbers. The maximum gain is recorded at zero spanwise wavenumbers at $St \approx 3.4$, indicating that the 2-D mechanism predominantly drives the flow. A rank-1 behavior emerges within the frequency-wavenumber range $3 \lesssim St \lesssim 4$ and $\beta/\beta_1 \leq 8$, where $\sigma_1 \gg \sigma_2$. Modal stress analysis reveals that each shear layer region is receptive to distinct frequency-wavenumber pairs. Therefore, actuators should be designed with specific objectives to effectively target particular shear-layer regions.

Comparing the resolvent model to SPOD reveals that the resolvent model captures a high-amplification energy frequency similar to the SPOD spectrum. The resolvent model predicts optimal amplification in the LSL at the dominant frequency, while coherent structures appear in the SPSL in SPOD. This discrepancy arises from a modeling assumption in the resolvent formulation, where the leading forcing is arbitrarily selected from broadband forcing and may not fully describe the flow. In our complex multi-stream jet flow, where multiple shear layers are present, the linear operator tends to over-optimize the amplification mechanism. Consequently, the sub-optimal modes align more closely with the leading SPOD mode. We speculate that resolvent formulations with the forcing models perhaps can potentially improve the linear prediction in such complex flows. We leave this as a direction for future work. Alternatively, optimizing over a subspace of the resolvent operator by restricting the input/output could better predict the dominant mechanism, as demonstrated in our component-wise input-output analysis.

To tailor practical control parameters, componentwise input-output analyses are conducted with state restrictions. These analyses reveal that velocity-based control strategies exhibit higher energy amplifications than their thermal counterparts. Among the cases considered, the 2-D streamwise velocity input produces the maximum energy amplification. A shift among the first three gain curves indicates a change in the optimal input-output location across the three main shear layers. Similar to SPOD analysis, the optimal mode resides in a single shear layer when the singular values are well separated. Further analysis incorporates spatial constraints on the input at the splitter plate trailing edge. For the input constraint case, the gain behavior deviates from cases without spatial restrictions. The higher gain observed at a higher 3-D wavenumber suggests that 3-D input at the SPTE results in the maximum flow response in the SPSL region at the dominant frequency. The insights from this study will guide the design of active control strategies for ongoing numerical simulations and experimental investigations.

\section*{Acknowledgments}

This material is based upon work supported by the Air Force Office of Scientific Research (AFOSR) under award number FA9550-23-1-0019 (PO: Dr.~Gregg Abate). We also acknowledge the Research Computing Center at Syracuse University for providing computational resources. The authors would like to thank Dr.~Mark N.~Glauser and Dr.~Fernando Zigunov for fruitful discussions.

\section*{Appendix}

\subsection{Reynolds number effect} \label{sec:Re_effect}

The mean flow exhibits the production and reflection of shocks, which combine to form a shock train in the flowfield (see figure~\ref{fig:jet_mean}). To accurately capture the correct mode shape, a drastic grid refinement is required when performing the resolvent analysis at a high Reynolds number of $Re = 2.7 \times 10^6$. Previous studies have reported that the grid resolution required to compute the correct forcing mode is even higher than the grid used for LES at the same Reynolds number~\cite{mckeon2010}.

For three different Reynolds numbers, differing by an order of magnitude, the resolvent gain is shown in figure~\ref{fig:Re_effect}(a). We observe that for the high Reynolds number of $Re = 1 \times 10^6$, few peaks are present in the gain curve, whereas for $Re = 1 \times 10^5$ and $8 \times 10^4$, these peaks smear out. As shown in figure~\ref{fig:Re_effect}(b), a noise-type structure is observed downstream of the aft deck at the high Reynolds number $Re = 1 \times 10^6$. We note that further refinement of the grid does not lead to a cleaner forcing mode. The response modes at all three Reynolds numbers exhibit no qualitative discrepancies. The Reynolds number can be considered a free parameter, as the amplification mechanism remains unchanged with varying Reynolds numbers~\cite{schmidt2018}. Our current work focuses on inviscid instability, which allows us to slightly lower the Reynolds number~\cite{pickering2021, doshi2022}. Consequently, we maintain $Re = 1 \times 10^5$ for our linear analysis, as the gain and modal structures remain consistent across all Reynolds numbers.
\begin{figure} [hbpt]
     \centering
         \includegraphics[width=1\textwidth]{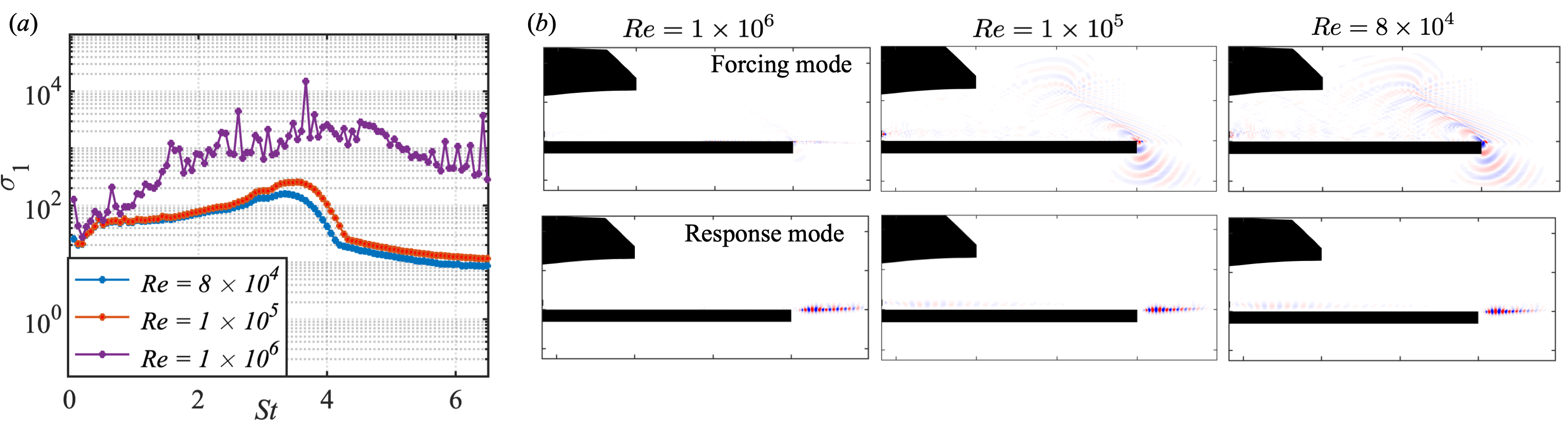}
    \caption{(a) The optimal singular value $\sigma_1$ at different Reynolds numbers ($Re$). (b) The forcing and response mode shape at $St =3.27$ and $\beta=0$ at the various $Re$. Forcing and Response: [blue, red] = [-2, 2].}
    \label{fig:Re_effect}
\end{figure}

\subsection{Randomised singular value decomposition} \label{sec:rsvd}

In fluid flow, the size of the linear operator becomes enormous for high Reynolds numbers, making the computation of the SVD of such a large matrix tremendously challenging due to significant memory and computational requirements. To address this, we use randomized singular value decomposition (R-SVD) to compute the singular vectors and singular values, as described by \cite{ribeiro2020}. The details of the algorithm are omitted for brevity, and the results from R-SVD and full SVD (Arnoldi method, MATLAB \textit{svds}) are compared in figure~\ref{fig:svd_vs_rsvd}. The leading singular value ($\sigma_1$) and mode structures show good agreement for various spanwise wavenumbers. Thus, we employ R-SVD to perform the singular value decomposition of the resolvent operator in this work.
\begin{figure} [hbpt]
     \centering
         \includegraphics[width=0.3\textwidth]{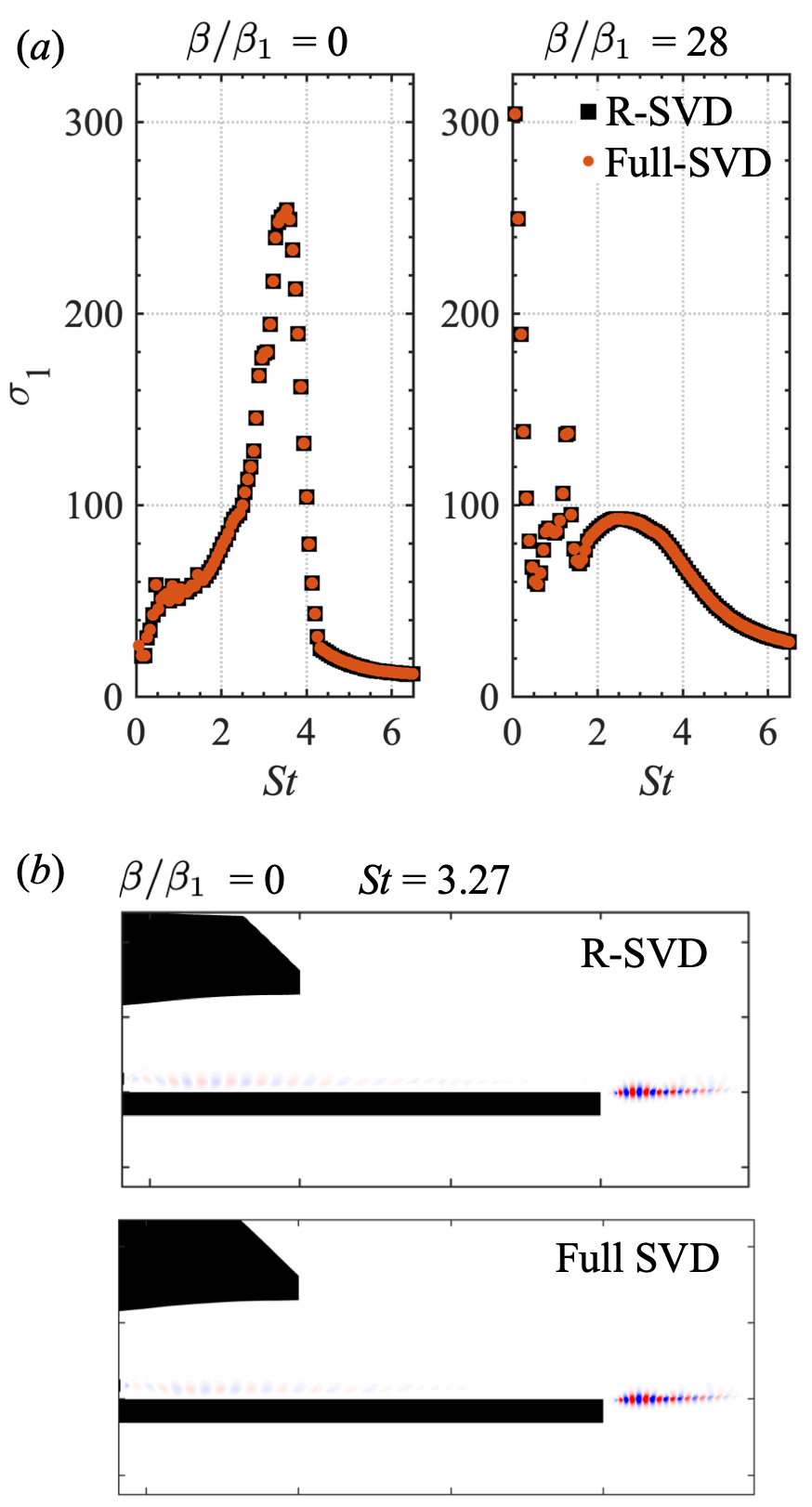}
    \caption{(a) The comparison of (a) the optimal singular value $\sigma_1$ and (b) response mode (real component of $v-$velocity) for randomized singular value decomposition (R-SVD) and full-SVD at $\beta/\beta_1=0$ and 28 with $\alpha D_h/\overline{u}_j = 0$. [blue, red] = [-2, 2]. }
    \label{fig:svd_vs_rsvd}
\end{figure}

\bibliographystyle{plain}
\bibliography{references}

\end{document}